\documentclass[sigconf,timestamp,nonacm,screen]{acmart}

\usepackage{hyperref}
\usepackage{graphicx}
\graphicspath{ {./figures/} }
\usepackage{algorithm}
\usepackage{algorithmic}
\usepackage{amsmath}

\AtBeginDocument{%
  \providecommand\BibTeX{{%
    \normalfont B\kern-0.5em{\scshape i\kern-0.25em b}\kern-0.8em\TeX}}}

\setcopyright{none}
\begin{document}

\title[Location inference on social media data for agile monitoring of public health crises]{Location inference on social media data for agile monitoring of public health crises: An application to opioid use and abuse during the Covid-19 pandemic}

\author{Angela E. Kilby}
\email{a.kilby@northeastern.edu}
\affiliation{%
  \institution{Northeastern University}
  \city{Boston}
  \state{Massachusetts}
  \country{USA}
}

\author{Charlie Denhart}
\email{denhart.c@northeastern.edu}
\affiliation{%
  \institution{Northeastern University}
  \city{Boston}
  \state{Massachusetts}
  \country{USA}
}

\renewcommand{\shortauthors}{Kilby and Denhart, 2021}

\begin{abstract}
The Covid-19 pandemic has intersected with the opioid epidemic to create a unique public health crisis, with the health and economic consequences of the virus and associated lockdowns compounding pre-existing social and economic stressors associated with rising opioid and heroin use and abuse. In order to better understand these interlocking crises, we use social media data to extract qualitative and quantitative insights on the experiences of opioid users during the Covid-19 pandemic. In particular, we use an unsupervised learning approach to create a rich geolocated data source for public health surveillance and analysis. To do this we first infer the location of 26,000 Reddit users that participate in opiate-related sub-communities (subreddits) by combining named entity recognition, geocoding, density-based clustering, and heuristic methods. Our strategy achieves 63 percent accuracy at state-level location inference on a manually-annotated reference dataset. We then leverage the geospatial nature of our user cohort to answer policy-relevant questions about the impact of varying state-level policy approaches that balance economic versus health concerns during Covid-19. We find that state government strategies that prioritized economic reopening over curtailing the spread of the virus created a markedly different environment and outcomes for opioid users. Our results demonstrate that geospatial social media data can be used for agile monitoring of complex public health crises.
\end{abstract}

\ccsdesc[500]{Applied computing~Health informatics}
\ccsdesc[300]{Applied computing~Economics}
\ccsdesc[500]{Human-centered computing~Social media}
\ccsdesc[300]{Information systems~Information retrieval}

\keywords{social media, text mining, natural language processing, opioid use disorder, drug abuse, opioid crisis, health informatics, public policy, public health, Reddit, computational social science, location inference}


\maketitle

\section{Introduction}

The opioid epidemic is a persistent public health crisis in the United States. In 2019, an estimated 10.1 million people misused opioids, and 49,860 died due to opioid overdose \cite{substance_abuse_and_mental_health_services_administration_key_2020,mattson_trends_2021} --  a more than sixfold increase since 1999. While the crisis has improved somewhat in recent years, with yearly deaths due to opioid overdose leveling off and even slightly declining beginning in 2017, emerging qualitative and quantitative evidence suggests that the Covid-19 pandemic may have intersected with the opioid epidemic, increasing strain on drug-using populations and contributing to a substantial reversal in these improvements in 2020. Preliminary CDC numbers show that opioid overdoses jumped by approximately 40\% from 2019 to 2020, representing the largest year-on-year increase since modern tracking of the crisis began in 1999 \cite{ahmad_provisional_2021}.

The causes of this dramatic increase in opioid overdose deaths during the Covid-19 pandemic are not yet understood \cite{scott_pandemic_2021,macmadu_comparison_2021}. A leading theory points to the economic and social disruptions to daily life due to the spread of the virus, which may have led to depression, anxiety, job loss, lack of social contact and therefore greater drug abuse. These disruptions may have compounded pre-existing social and economic strain and malaise that has been associated with rising ``deaths of despair'' from drug overdose, suicide, and alcoholism among the American working class over the past two decades  \cite{case_rising_2015}, and may be caused by declining economic and social prospects for large swaths of American society \cite{case_mortality_2017}. However, other research has highlighted that the opioid crisis may be driven by supply factors, from the widespread availability of prescription opioids in the 2000-2010 era to the increase in availability and potency of illicit opiates starting in 2010 \cite{ruhm_deaths_2018}. Key data points do not align with a theory of economic and social stress being the primary driver of increased overdose mortality during the pandemic, including preliminary CDC statistics on suicides, which show a 5.6\% decline between 2019 and 2020 \cite{ahmad_leading_2021}. This remarkable drop in suicides suggests the possibility of \emph{reduced} economic strain, perhaps arising from expansions in the social safety net in response to the Covid-19 crisis.

In this paper, we introduce a rich, geolocated dataset for public health surveillance and policy study. Data were extracted from Reddit, a social media and content aggregation website that is divided into a large number of sub-communities, called \emph{subreddits.}  We focus on the Covid-19 pandemic and its intersection with the opioid epidemic, leveraging conversations to gain a quantitative and qualitative understanding of opioid users' experiences. Because we have engaged in a location inference procedure to produce a geographically tagged user cohort, we are able to answer pressing policy questions about the impact on wellbeing of local policies balancing competing concerns of economic reopening and viral spread. Such geographic social media data sources may also be used to answer other policy questions about the opioid crisis.  How do prices change after law enforcement actions? Do locations of exchange between buyers and sellers, such as open-air drug markets, change \cite{whelan_even_2020}? Does demand shift between different types of opioids? Are supply chains affected and if so, how?  Are objectives of policy implementations in fact met?  With richer data on opioid black-markets, these questions can be investigated more scientifically. 

Since transaction data is not recorded for illicit markets, as it would be with prescription data, data must be collected by self-reported means such as surveys or crowdsourcing (which are usually anonymous) \cite{dasgupta_crowdsourcing_2013}.  We choose to utilize data that is self-reported conversationally on Reddit.  Using this data we attempt to infer the locations of users in drug related subreddits, which will allow for location-aware analyses of the impact of opioid policies and other changes in drug supply. 

\section{Related Work}
\label{litreview}

The two bodies of work most relevant for this project are those that perform location inference tasks and those that analyze the behavior of black-market opiates users. 

\subsection{Location Inference using Social Media Data}

Twitter has historically been the primary data source for location inference tasks.  This is largely because a small sample of tweets are tagged by the tweeter with their posting location, allowing for the possibility of directly supervised learning. Several types of data such as tweet content, user metadata, and social networks of users have been utilized to infer location, and a variety of machine learning techniques have been employed including named entity recognition, gazetteer matching, probabilistic modeling, and graph-based modeling \cite{ajao_survey_2015}. Location tagging on Twitter has specifically been used to assess misuse of prescription opioids across locations. Opioid conversation locations were compared to distributions of drug users provided by the 2013–2015 National Surveys on Drug Usage and Health, finding high correlations in estimates of geographic prevalence between Twitter data and gold-standard epidemiological survey data \cite{chary_epidemiology_2017}.      

There is, additionally, a growing body of work that leverages Reddit. Previous identification of users on Reddit has been conducted using document ranking and query expansion strategies \cite{balsamo_firsthand_2019}.  Assigning location to users on Reddit has been performed primarily by using posts in which users explicitly declare where they live (e.g. 'I live in ...') \cite{harrigian_geocoding_2018,balsamo_firsthand_2019}. Hand labeling of data subsets by trained annotators has also been exercised to evaluate location inference algorithms \cite{harrigian_geocoding_2018}. Our methodology overlaps with some of the approaches used in these studies such as the use of a Gazetteer (Geonames) and an out-of-the-box named entity recognition tool (Spacy), as well as a trained annotator to assess model accuracy.

Data collected from the web and social media have also shown to be useful for monitoring different infectious diseases.  Voluntary participation in mobile applications for online monitoring of the H1N1 pandemic has been explored \cite{freifeld_participatory_2010}.  Additionally, informal sources, such as Twitter and HealthMap, were leveraged to track the geographic spread of the 2010 Haitian cholera outbreak \cite{chunara_social_2012}.

\subsection{Behavior of Black-market Opioid Users}
Many sources of black-market behavior stem from survey and crowdsourcing \cite{dasgupta_crowdsourcing_2013}.  These are valuable sources of information and allow for direct access to rich street level behavior.  They are, however, not very scalable as surveys are tedious to carry out and few users are willing to report their behavior.  Therefore, it is difficult to gain representative insight into black-markets as a whole.  There is some recent work, however, that has leveraged geocoded users from Reddit in order to assess user prevalence across states \cite{balsamo_firsthand_2019}.  This approach was able to locate a distribution of users in the United States that corresponds to United States Census data.  Topic modeling has been applied to drug related conversation to identify major themes \cite{pandrekar_social_2018}. There has also been research into the intersection of opioid use and the coronavirus pandemic in the context of social media, such as findings of increased discussion of major opioid use disorder themes during the pandemic \cite{krawczyk_how_2021}.  Shifts in social network behavior and community support have also been found in response to the pandemic \cite{bunting_socially-supportive_2021}.

\section{Data}
\label{secdata}

Reddit can be thought of as a collection of topics (subreddits) that each contain posts by users, organized as trees.  Posts at the top-level of the tree are considered submissions and contain a title, while all responses to a post are considered comments.  Other than differences in the position in the posting tree and the presence (or absence) of a title, in our location-identification exercise, we do not distinguish between submissions and comments, nor do we utilize the tree structure of submissions and comments.  

The \href{https://praw.readthedocs.io/en/stable}{Python Reddit API Wrapper (PRAW)} and the \href{https://psaw.readthedocs.io/en/latest}{Pushshift API (PSAW)} were used to request data from Reddit. Data collection started in 2016 and has continued periodically until the writing of this paper. 
Our dataset for location inference was constructed in two steps.  First we crawled the subreddit \href{https://www.reddit.com/r/opiates/}{r/opiates} and \href{https://www.reddit.com/r/heroin/}{r/heroin}, collecting all posts from 2007-2021. These two subreddits are the primary communities on Reddit for black-market opiate use discussion. We then collected the entire posting history across all of Reddit for each user who participated in r/opiates and r/heroin discussions. In total, we collected 91,095,755 posts (85,639,422 comments and 5,456,333 submissions) from 116,661 distinct users.  These posts were derived from 298,401 distinct subreddits between November 2005 through September 2021. Our dataset for policy analysis was a subset of the dataset constructed for location inference, comprising only the subset of data in larger dataset that was posted to r/opiates and r/heroin. 

\section{Methodology}

\subsection{Pre-processing}

The most relevant information contained within all of the submissions and comments, for our purposes, are location entities.  In order to extract these, we use the named entity recognition tagger from Spacy \cite{montani_explosionspacy_2021}.  In particular, we use those entities labeled as `GPE' (geopolitical entity) by Spacy.  Since we did not hand label geopolitical entities in the context of Reddit (or any corpus containing similar language), we did not fine-tune Spacy's named entity recognition tagger.  The out-of-the box tagger performs sufficiently, but seems to struggle more with the distinct dialogue of Reddit than it does with more formal language.  Therefore, we filter the `GPE' entities produced by Spacy. We remove any entities that exist in a hard-coded set of entities that we observed and decided would not be useful [see Appendix \ref{GPEfilter}]. We also remove entities that do not exist in a gazetteer of known global locations, Geonames\footnote{\url{https://www.geonames.org/}}.  We also extract entities via the name of the subreddit that a submission or comment is posted in, including the subreddit name as an entity if it corresponds to a location. We obtained a list of location-based subreddits from a Reddit page.\footnote{ \url{https://www.reddit.com/r/LocationReddits/wiki/index}}  All entities are converted to lower case across all data sources used.

In order to make the extracted entities more uniform and utilizable for the geocoder, we also apply several transformations to eligible entities.  If an entity matches a state abbreviation (from the United States), we expand the entity to the full name of the state. We also utilize a hard-coded list of location `nicknames' to convert an entity to its formal name (e.g. `Vegas' to `Las Vegas') [Appendix \ref{aliases}].  Finally, for large U.S. states, we represent the entity as several GPEs mapped to different regions across the state using a hard-coded map of state to regions (e.g. `California' would become `Southern California' and `Northern California') [Appendix \ref{state}]. This is useful for the clustering portion of the process (described below) because it allows geographic clusters that may exist within a large state to leverage keywords mentions of the state name. Since a state only geocodes to a single latitude/longitude pair, a cluster containing that pair would otherwise be unfairly weighted relative to other possible clusters existing within a large state.

\subsection{Extracting User Locations}

In order to disambiguate the possible real locations that can be associated with a textual entity, we use the Geonames geocoding service.  For each entity, we retrieve all compatible geocodes as determined by the Geonames database.  We pool all compatible geocodes from each entity together, then cluster all of the possible latitude and longitude pairs retrieved from the geocodes using DBSCAN\footnote{\url{https://scikit-learn.org/stable/modules/generated/sklearn.cluster.DBSCAN.html}}.  We elected to use DBSCAN because of the close alignment between density-based clustering and the domain problem.  Two parameters are required in DBSCAN.  The first parameter allows us to specify how far we want locations to be from each other in order to be considered part of the same cluster.  We use euclidean distance to measure distance between latitude/longitude pairs.  We found that a euclidean distance value of 2.5 roughly corresponds to 100 miles of separation between points, which we found sufficiently distinguishes between clusters. The second parameter allows us to define the minimum number of points necessary to form a cluster.  We considered a single location in isolation to insufficiently represent a user's location, so we use a minimum cluster size of 2. 

Once we have a collection of clusters, each containing 2 or more locations, we identify a representative member of each cluster. This allows us to compare location clusters of a user more easily and to assign a single location to a user once we have identified the most likely cluster that the user belongs to. For example, if a user has mentioned several locations in Massachusetts that belong to the same cluster, we want to identify a single one of those locations to represent the cluster.  In order to select this representative member, we use the following comparison rules to sort locations in a cluster such that more likely representations of a the cluster are ordered ahead of less likely ones.

\begin{enumerate}
    \item the most granular location (e.g. a location with a city, state, and country is prioritized over a location with just a state and country)
    \item the location most frequently mentioned by the user
    \item the location with the largest population
\end{enumerate}

In our Massachusetts example, if a user mentioned the words `Massachusetts', `Boston', and `Lowell' each once, `Boston' and `Lowell' would be sorted ahead of `Massachusetts' because they are more granular.  `Boston' and `Lowell' have been mentioned each just once, but since `Boston' has a greater population, it then gets sorted ahead of `Lowell'.  We take the first item from this sorted list of locations and assign it to be a representative member of the cluster.  In our Massachusetts example, that would be `Boston'.

Finally, we assign a score to the representative member of each cluster using our scoring process described in Section \ref{confscore}. The representative member with the highest score is considered to be the inferred location for that user.  

\begin{algorithm}
\caption{Ranking user location entities}

\begin{algorithmic} 
\REQUIRE User $U$

\STATE $entities \leftarrow \text{extract\_entities}(U)$

\FOR{$e$ in $entities$}
\STATE geocode $e$
\STATE extract coordinates from geocodes
\ENDFOR

\STATE $clusters \leftarrow$ cluster coordinates

\FOR{$c$ in $clusters$}
\STATE $guess \leftarrow$ best geocode in cluster using $priority(cluster)$
\STATE $score \leftarrow$ combine guess features with $score(guess, U, c)$
\ENDFOR

\STATE Return guesses, scores

\end{algorithmic}
\end{algorithm}

\subsection{Labeling}
A trained annotator was employed to inspect the full posting history of 800 random users in our cohort and ascertain their locations and associated timeframes.  We use notes and results of 600 of these users to fine-tune heuristics and hard-coded values in our algorithm.  We consider the remaining 200 users to be a hold-out test set in which performance values are most honest, since the results of these users were not utilized in the construction of the algorithm.

\subsection{Confidence Scoring}
\label{confscore}
We also create a confidence scoring method for the likelihood of successful location identification  by the location inference algorithm for each user.  In order to do this, we use the manually annotated data as labels, with a score of 0.0 assigned to a miss, 0.5 assigned to partially correct guess, and 1.0 assigned to a fully correct guess. We consider a partial guess to contain the correct state but not the correct city. We then train a regression model that predicts a correct location guess versus a miss using a collection of features that represent key components of the dataset used for the location inference process.  The features we utilize include: 
\begin{enumerate}
    \item the percentage size of the cluster that the guess belongs to out of to all guesses for the user
    \item the total number of entities extracted for the user
    \item if the location is in the United States
    \item the total number of posts by the user
    \item the duration of the posting history for the user on Reddit
    \item the population of the location guess
\end{enumerate} 

We train two random forest regression models to assign a confidence score given the feature set described above for each user. The first model, the \textbf{positive model}, assigns a confidence score for users that had a location guess and the second, the \textbf{negative model} for users without a location guess.  Two different models are necessary because different feature sets are available for the two options. In particular, the second model does not contain, the percentage size of the cluster that the guess belongs to out of to all guesses for the user, if the location is in the United States, or the population of the location guess.

The regression model assigns weights to the importance of each feature in predicting whether a location guess is right or wrong. For example, for the positive model, the random forest regression model identified the duration of the posting history for the user on Reddit as the most important to a successful location guess. All other features (except whether the location was in the United States, which was not very important) were approximately equally as important. These trained regression models are then used to predict the chance that a user location guess is correct, and thus produces a confidence score for each user's location.

This confidence score critically allows us to select a digital cohort for whom we have a high confidence that our location prediction is correct.

\section{Results}

\subsection{Performance}

\subsubsection{Location Inference}

To evaluate the location inference model, we consider a location correct if the algorithm's best guess matches the hand annotator's location at the city-level for users in the United States, at the country-level for international users, or if it correctly does not produce a location for a user for which a location could be found by the hand-annotator.  On the hold-out test set of 200 users, the algorithm correctly assigns a location to a user 59 percent of the time.  The algorithm further was able to partially assign the correct location to a user 4 percent of the time.  We consider a location partially correct if it is incorrect at the city-level for users in the United States, but correct at all other levels of granularity.  Combining fully correct guesses with partially correct guesses yields 63 percent accuracy at the state/country level. (Performance on the full 800 user hand-annotated sample was 68.7\%.)

\subsubsection{Confidence Scoring}

As described in Section \ref{confscore}, a scoring system was developed to numerically represent confidence in our location guess. This score is used in later sections to select a geographically-identified user cohort for which we have a high degree of certainty their location guess was correct.

From the full set of 800 manually labeled instances, we hold out 33 percent for testing and train our models on the rest.  We evaluate and present performance of the positive model, the a model for certainty of users that had location guesses. We present performance using a binary classifier for better interpretability of results, assuming fully correct and partially correct guesses to have a label of 1 and incorrect guesses to have a label of 0.  The positive model for users with a location guess achieves an AUC score of 0.71 and the negative model for users with no location guesses achieves an AUC of 0.64.  We then re-train this regression model on the full manually labeled data set and apply it to all users in the full sample, generating a score for each user in the study.

\begin{figure}[ht]
\includegraphics[width=0.9\columnwidth]{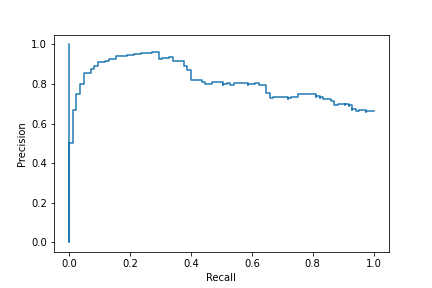}
\caption{Precision Recall Curve for Positive Model}
\end{figure}

\begin{figure}[ht]
\includegraphics[width=0.9\columnwidth]{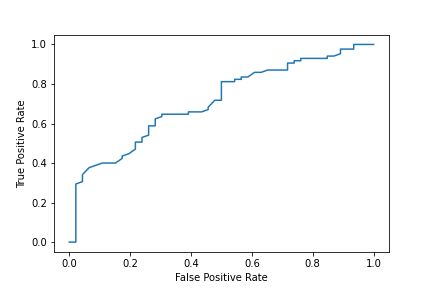}
\caption{ROC Curve for Positive Model}
\end{figure}

\begin{figure}[ht]
\includegraphics[width=0.9\columnwidth]{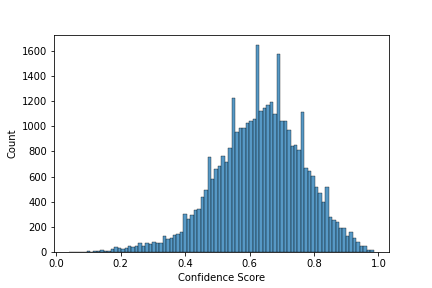}
\caption{Distribution of confidence scores}
\end{figure}

\subsection{User Characteristics}
Running the algorithm on our cohort of over 116k users resulted in cities for 23.8k users, states for 28.2k users, and countries for 37.9k users.  Roughly 60 percent of users, therefore, are deemed to not have a location that can be inferred. We filter out users with a confidence score of less than 50\%, suggesting that the remaining users are likely at least correct at the state/country level according to our confidence scoring process.  After filtering, the city, state, and country counts became 22.1k, 26.0k, and 33.4k respectively.  

\begin{figure*}[htbp]
  \centering
\includegraphics[width=0.8\textwidth]{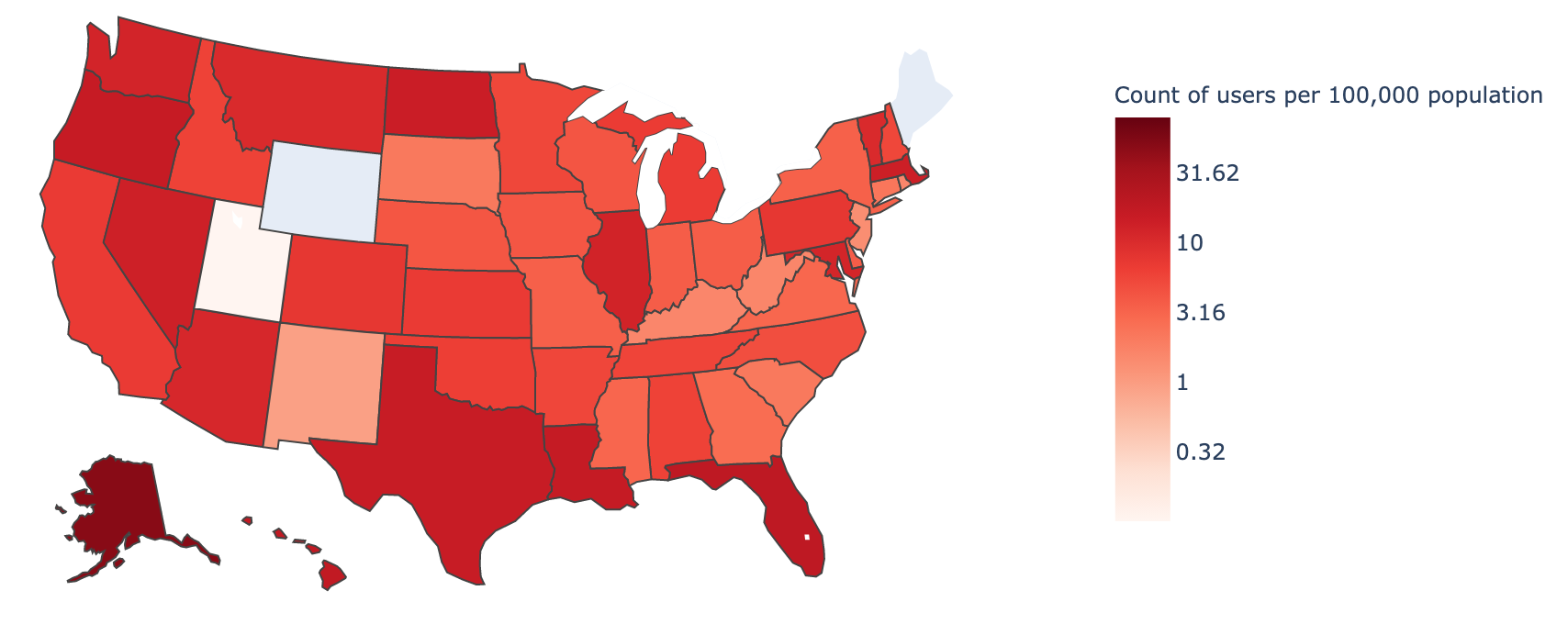}
\caption{Heatmap of Users by State: count of users in user cohort per 100,000 population}\label{heatmap}
\end{figure*}

Our cohort of users are from 328 distinct US cities, 48 US states (all states and D.C., except Wyoming and Maine), and 133 different countries. Figure \ref{heatmap} depicts their geographical distribution within the United States.

\subsection{Limitations of Approach}
Since our approach to location inference consists of many interdependent components, there are a few possible sources of error.

The mis-tagging of `GPE' entities by Spacy is one possible source of error.  There were approximately 150k distinct `GPE' entities across all posts in our database pre-filtering, but only around 5k post-filtering.  This suggests a high false-positive rate in `GPE' tagging.  The false-negative rate of `GPE' tagging (not tagging `GPE' entities that should have been tagged) is also a possible issue, but this has only been found observationally.  

Although our heuristic approaches were incorporated largely to combat sources of ambiguity, they also may introduce some error.  The geocoding of entities is a possible source of error as a textual entity could correspond to many places around the world.  

Finally, the largest threat to our approach is that it is context agnostic.  Any reference to a location by a user is considered equally (if our approach is able to detect it) regardless of whether the user is referring to their actual location or some arbitrary location.  In other words, our strategy rests on the assumption that a user's location can be inferred by the location that they most frequently discuss.  Our approach is also time agnostic, so our inference about a user's location refers to the location in which they spent the majority of their time over the course of their posting history on Reddit.

\section{Policy Application: The Opioid Epidemic and Covid-19}

We next utilize our geo-located user cohort to conduct a digital epidemiology study of black-market opioid user experiences during the Covid-19 pandemic. As described in Section \ref{litreview}, a small number of recent studies have used social media to understand the impact the Covid-19 pandemic has had on opioid users. However, our location-aware dataset allows us to answer richer health policy questions. Specifically, we will study the impact of state-level policies that prioritize re-opening the economy versus containing the spread of the virus, proxying for the degree to which state policymakers prioritized reopening the economy over curtailing viral spread with the state-level vote share for Donald Trump in the 2016 and 2020 presidential elections.

\subsection{Methodology}

\begin{figure}[ht]
\includegraphics[width=0.7\columnwidth]{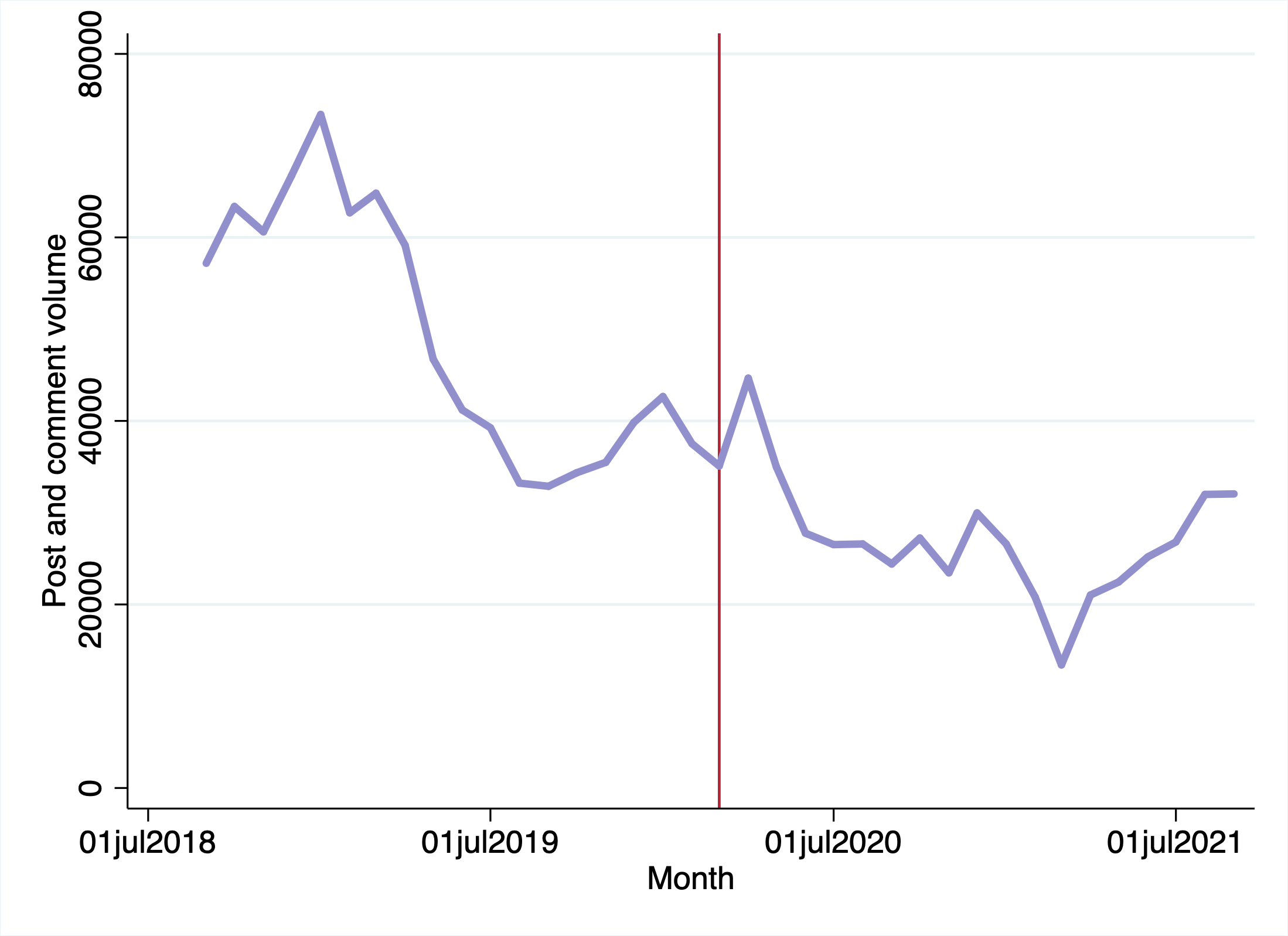}
\caption{Posting volume for period of study analysis, September 2018 to September 21; red line demarcates March 2020, the beginning of the pandemic}
\label{figvolume}
\end{figure}

Our dataset for the policy application is described in Section \ref{secdata} above, and comprises all user posts and comments in the subreddits r/opiates and r/heroin. We extract posts and comments for the time period between September 2018 and September 2021. Figure \ref{figvolume} depicts the overall post volume over time in the dataset for this period. During the study period, there has been declining overall engagement with r/opiates and r/heroin.

In order to represent key elements of the social and economic experience of black-market opiates users during the period before and after the pandemic, we develop \emph{topic}-\emph{keyword} mappings to represent key topic themes. Table \ref{topickeyword} displays the topic-keyword mappings. Counts of all keywords associated with each topic were extracted from our dataset, which is pre-processed by lemmatizing words, and an \emph{adjusted counts} metric was created by dividing the keyword count for each topic by the overall post volume of that month, the metric depicted in Figure \ref{figvolume}.

\begin{table}
  \caption{Topic-keyword mappings}
  \label{topickeyword}
  \begin{tabular}{cc}
    \toprule
    Topic & Keywords\\
    \midrule
    covid-19 & covid, virus, expose, pandemic, \\
             &  quarantine, corona, vaccination, \\ 
    crime    &  arrest, bust, narc, nark  \\
    drug     &  heroin, oxy, dope, fent, stimulant, \\            & diacetylmorphine  \\
government money & unemployment, irs, stimulus    \\
money        &  money, pay, spend, account, bill,\\ 
             & bank, broke, fund, payment, finance,\\
             & wage, salary, bankrupt, skint  \\
narcan       & narcan, naloxone   \\
overdose and death & die, overdose, death, dying, o.d. \\
physical     & pain, withdrawal, tolerance, addict, \\
             & sick, junkie, hurt, mental, health, \\
             & ill, hook, withdraw, puke, suicide, \\
             & vomit, nauseous, dopesick, junky \\
recovery prescriptions & methadone, suboxone, \\ 
             & buprenorphine, subutex   \\
  \bottomrule
\end{tabular}
\end{table}

\subsection{National Analysis of Black-market Opiate User Experiences during the Covid-19 Pandemic}

\begin{figure*}[htbp]
\includegraphics[width=0.24\textwidth]{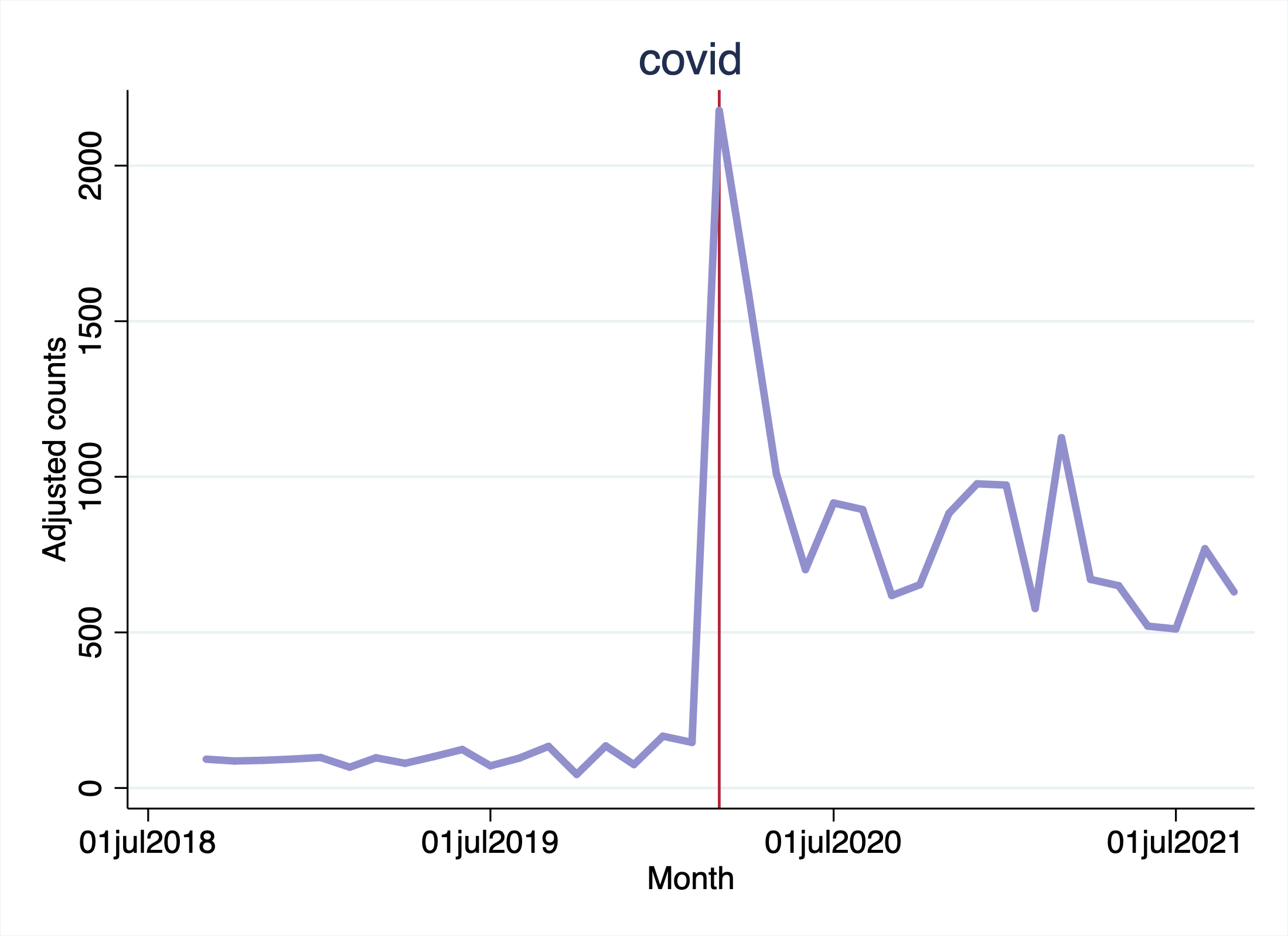}%
\includegraphics[width=0.24\textwidth]{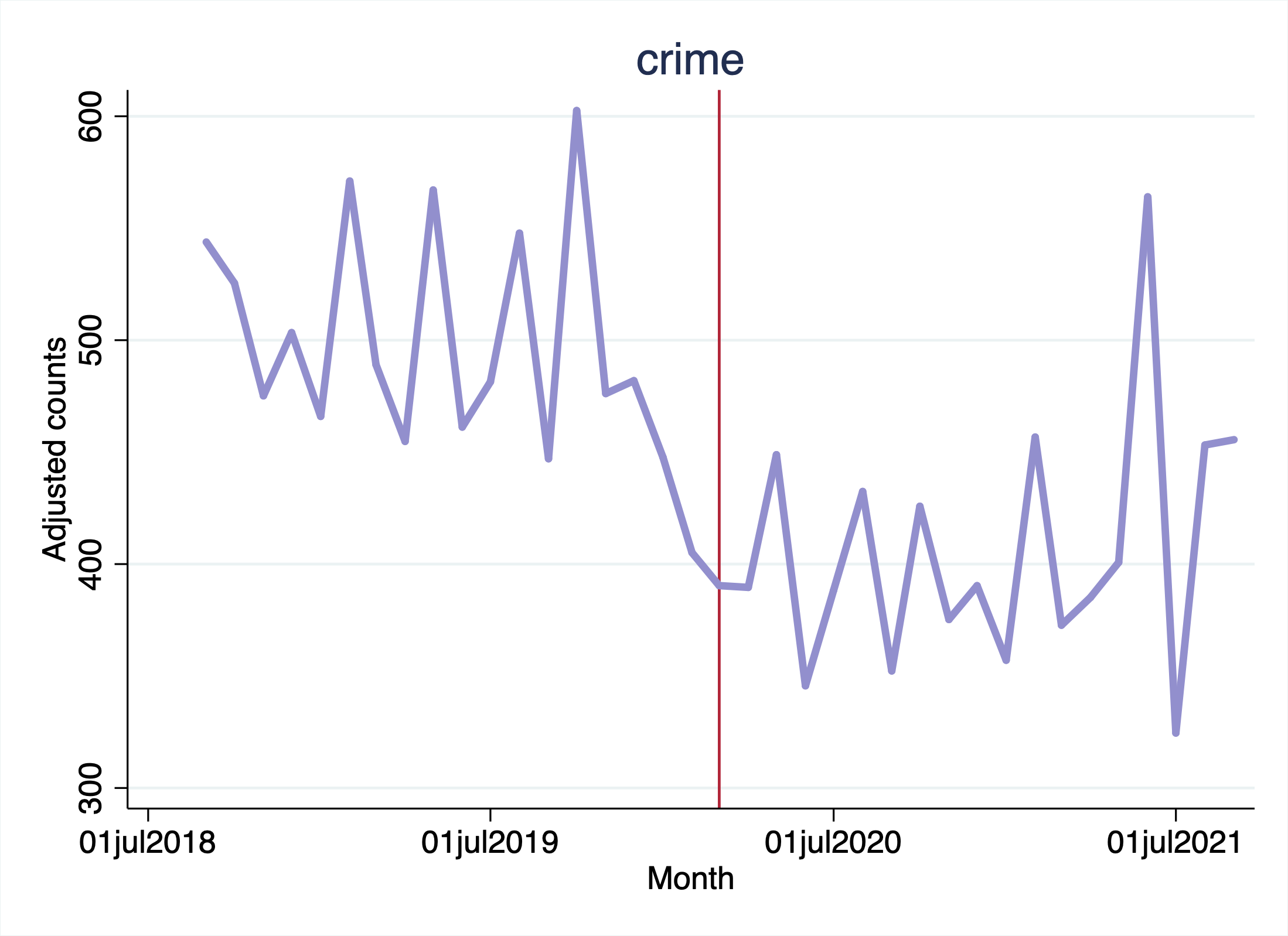}%
\includegraphics[width=0.24\textwidth]{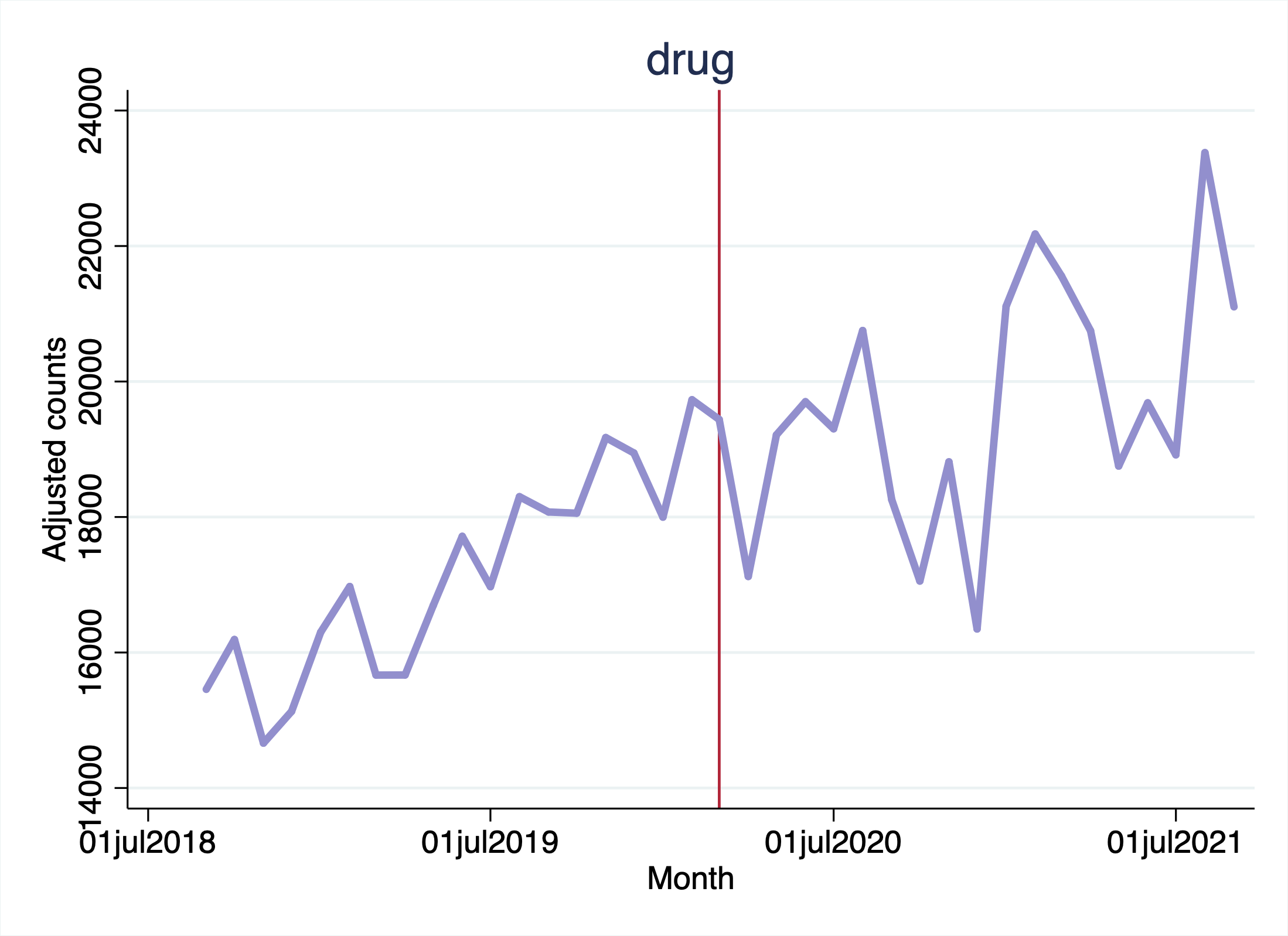}
\includegraphics[width=0.24\textwidth]{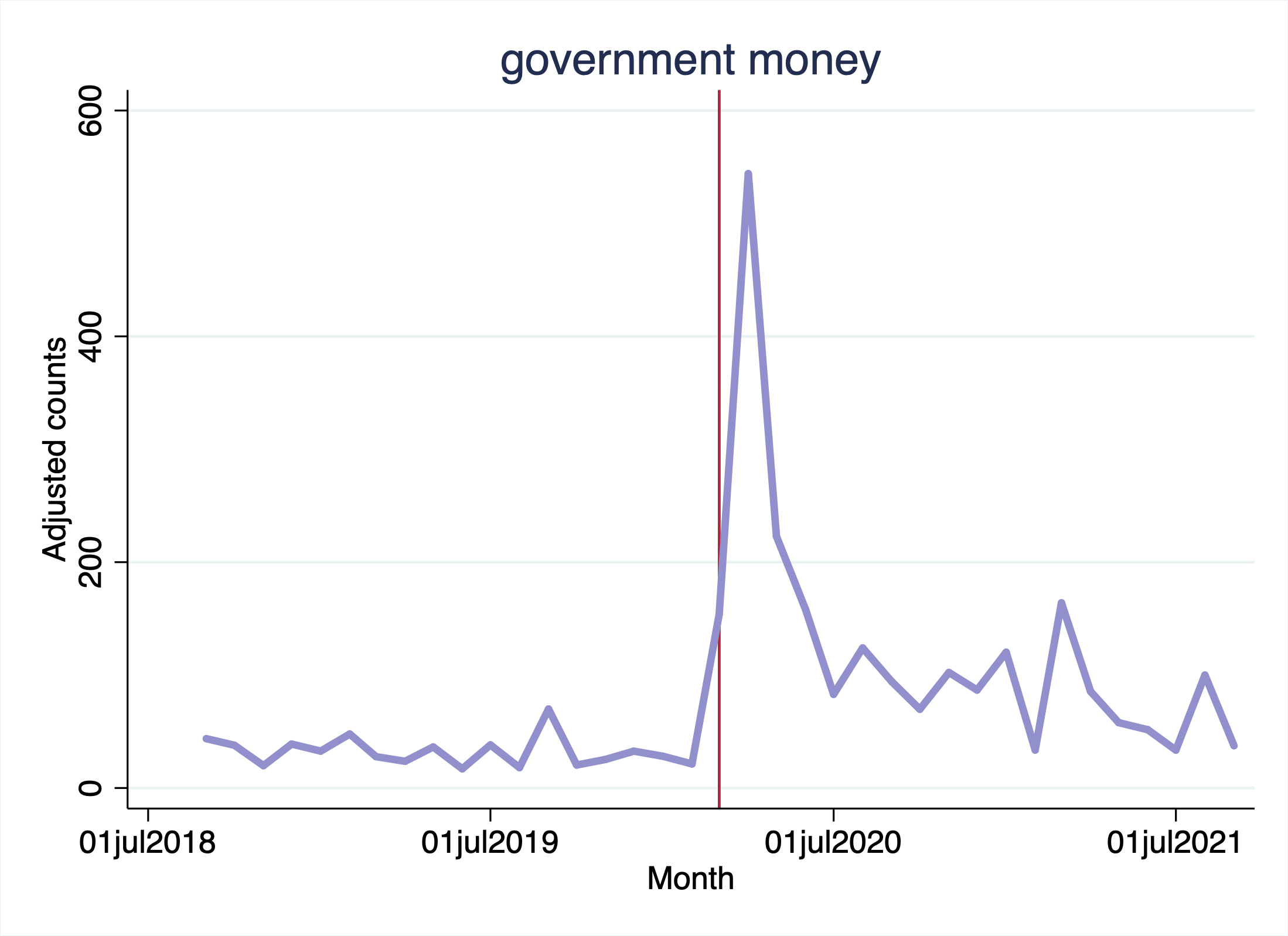}%
\includegraphics[width=0.24\textwidth]{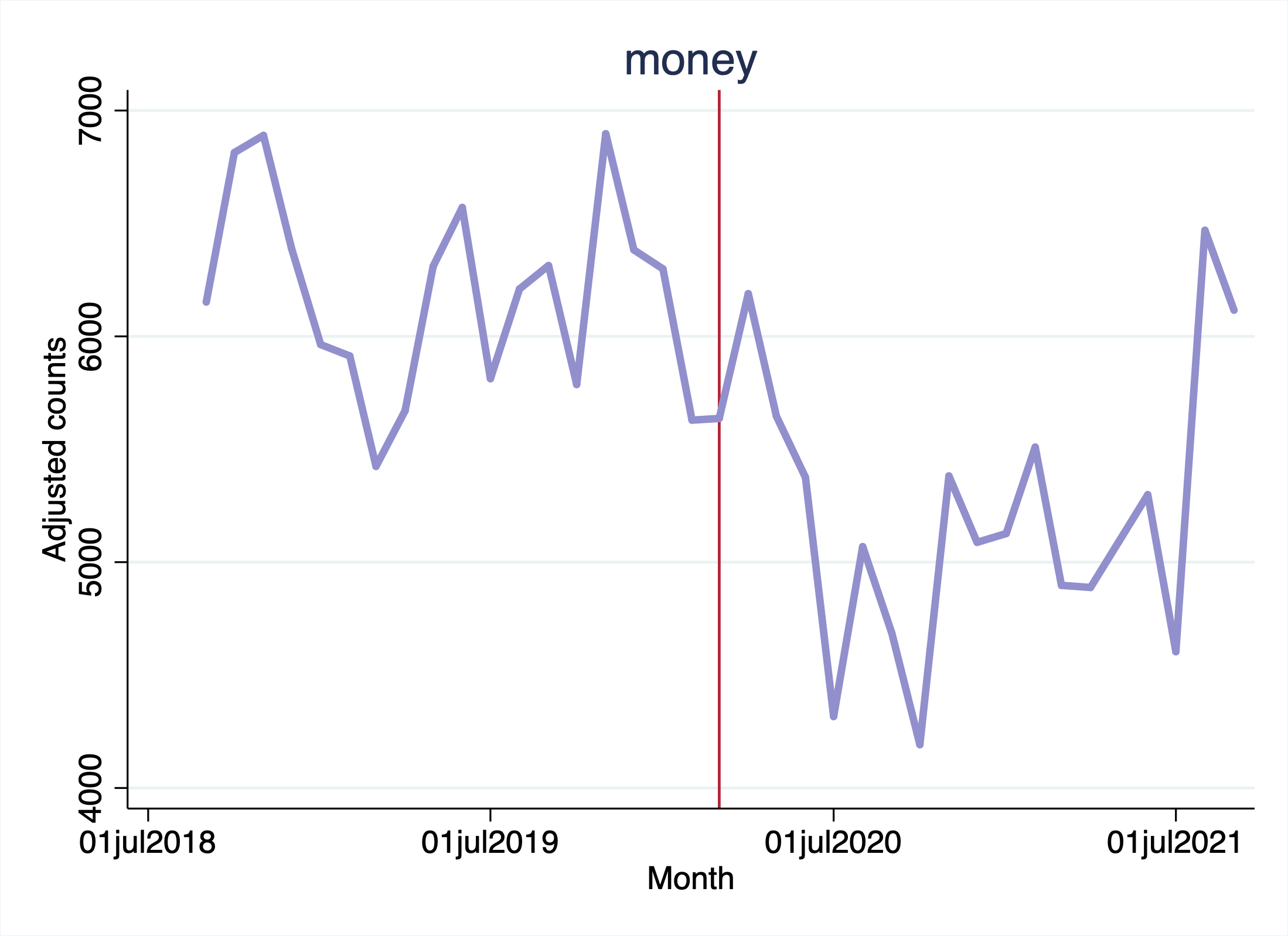}%
\includegraphics[width=0.24\textwidth]{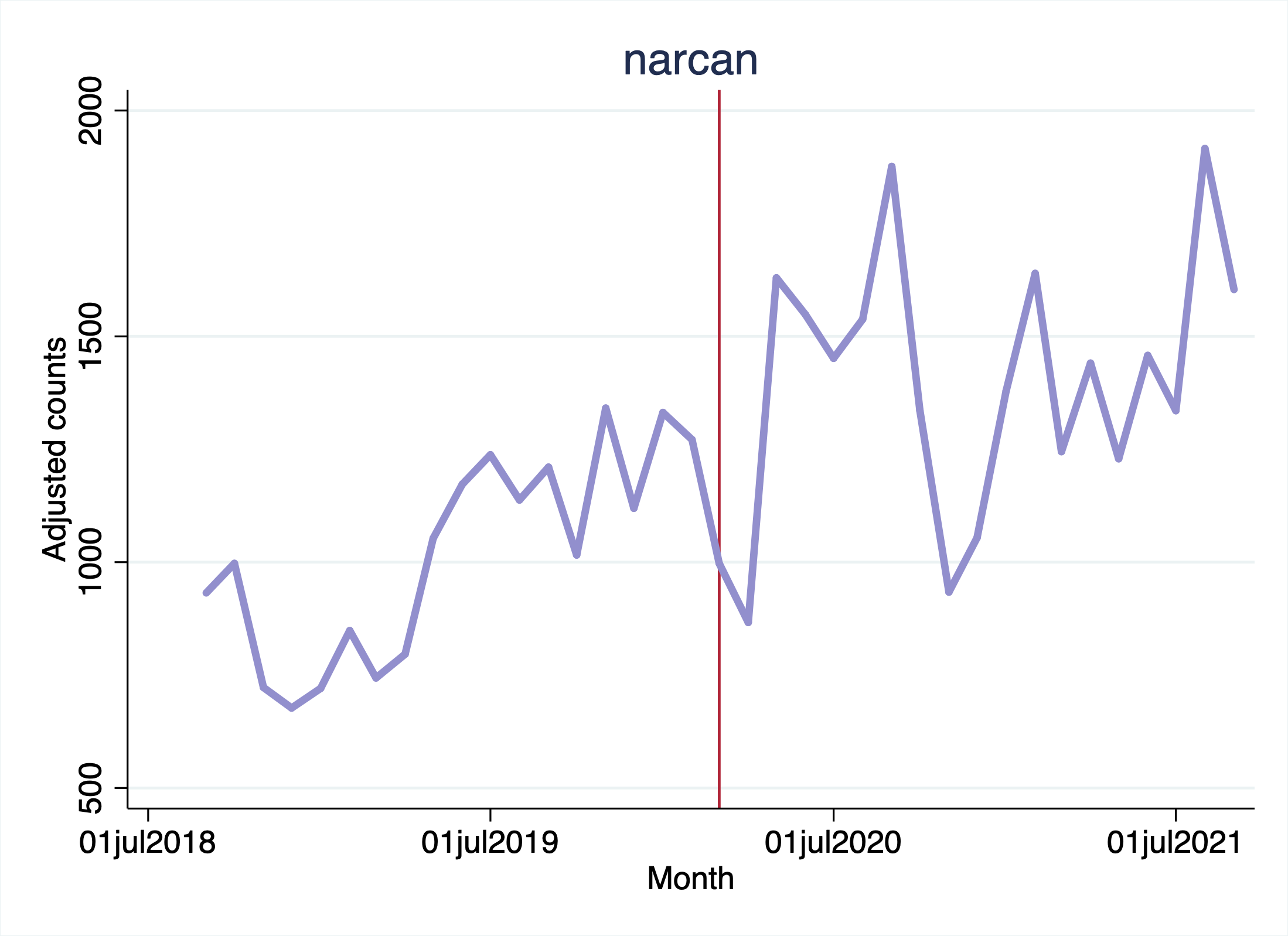}
\includegraphics[width=0.24\textwidth]{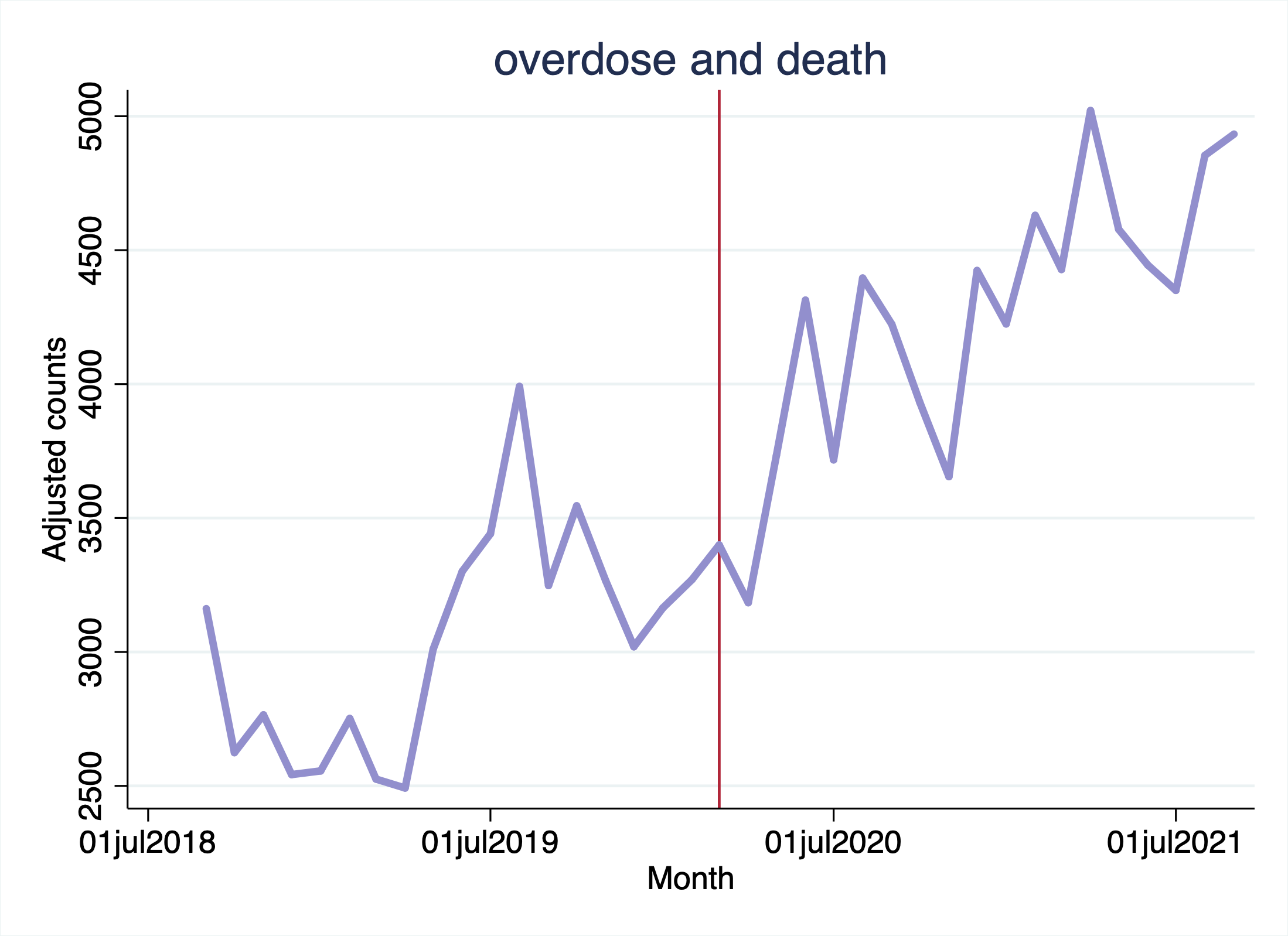}%
\includegraphics[width=0.24\textwidth]{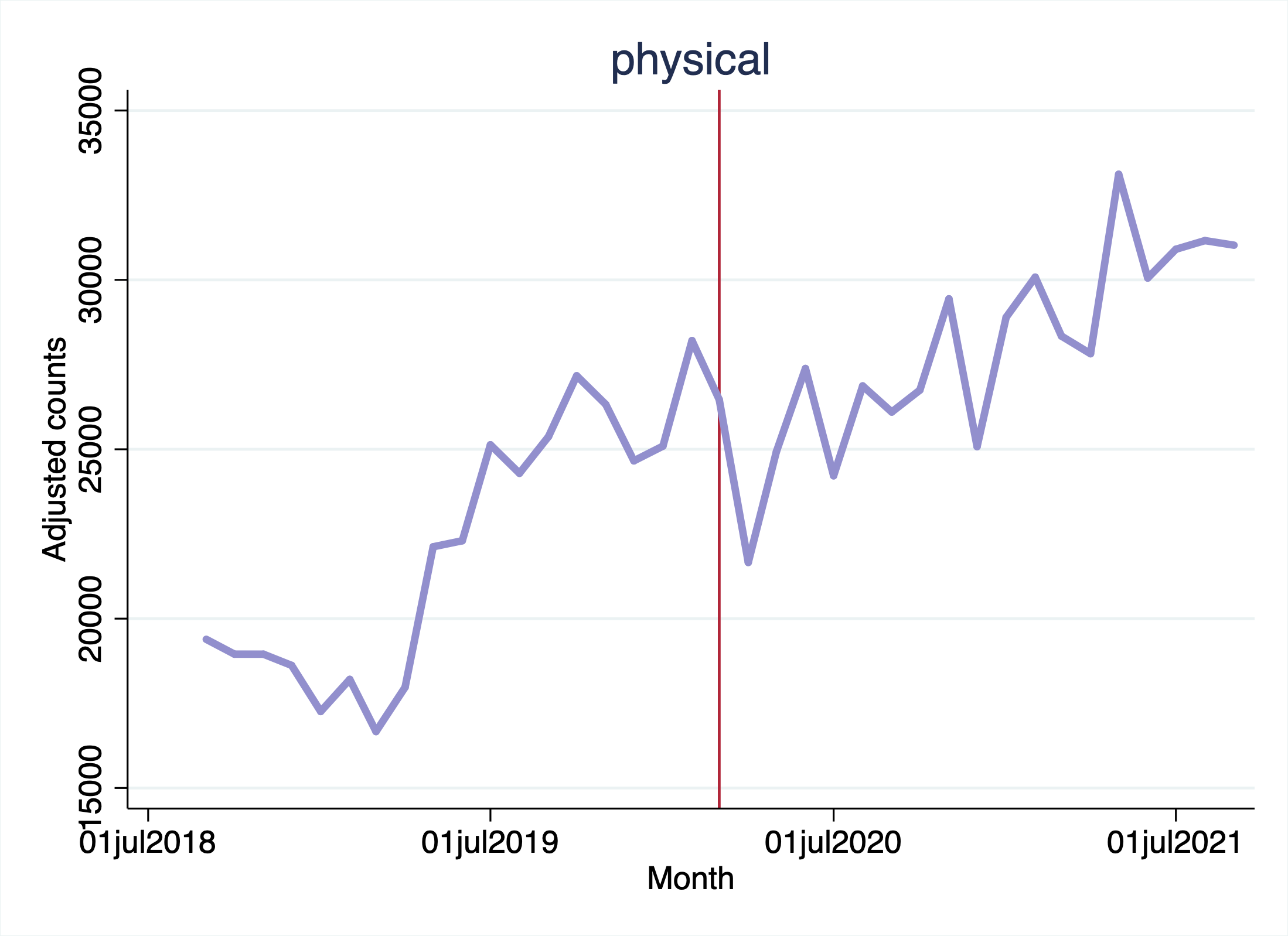}%
\includegraphics[width=0.24\textwidth]{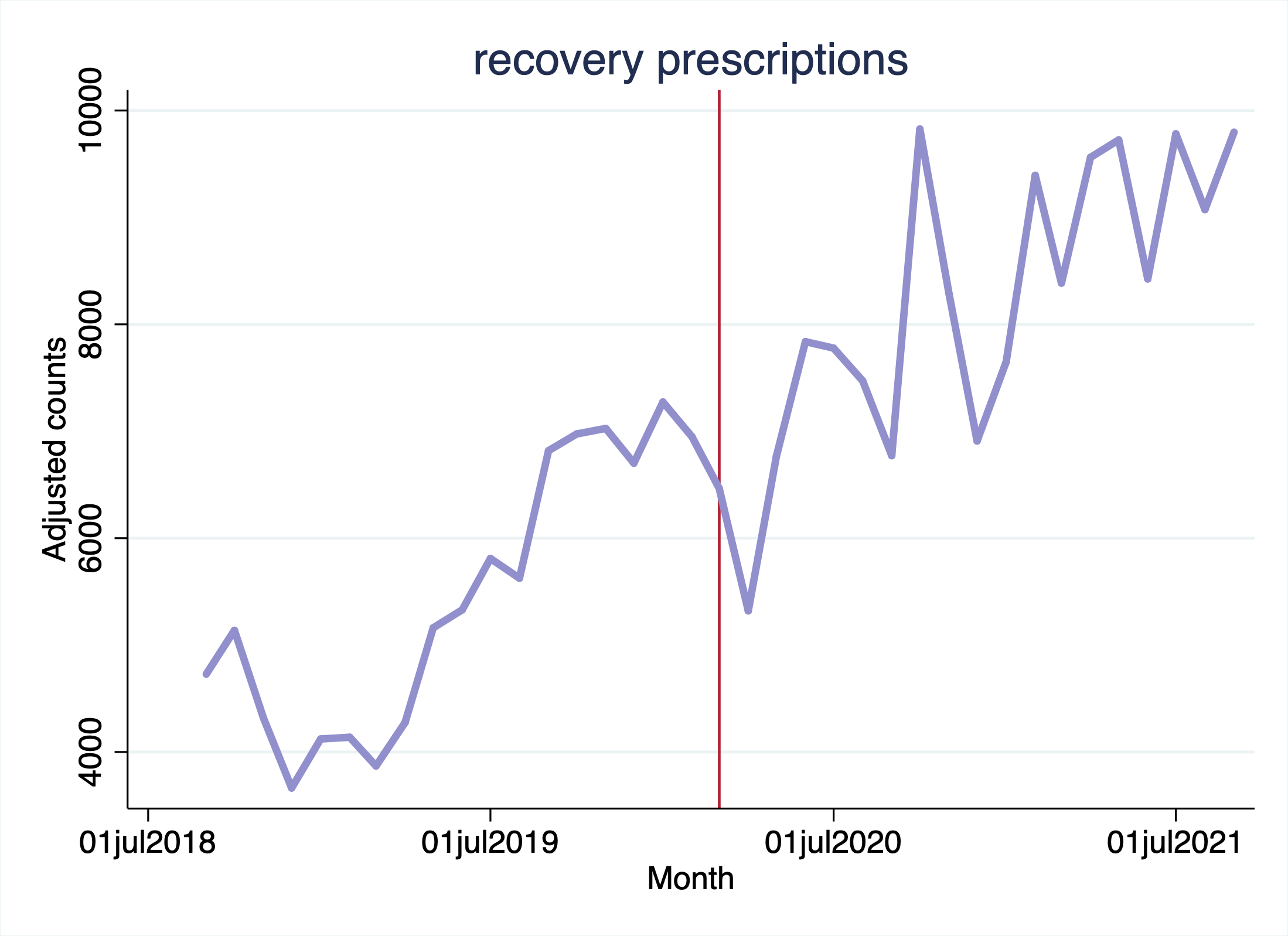}
\caption{Adjusted posting volume for period of study analysis, September 2018 to September 21, for each topic; vertical red lines demarcate March 2020, the beginning of the pandemic}
\label{figvolumekey}
\end{figure*}

We first present discussions by topic from the full dataset of r/opiates and r/heroin discussion. Figure \ref{figvolumekey} depicts adjusted post counts over time for each of the nine thematic topics presented in Table \ref{topickeyword}. There are several results of note. First, we can see that Covid-19 generated considerable discussion on the subreddit beginning in March 2020, and discussion continued throughout the pandemic. Second, there is some discussion of government relief programs - unemployment and stimulus checks - with spikes in April and December 2020, when stimulus checks were being distributed.

Discussions of drug use are trending upwards through the period before and after the pandemic began, but there is a notable spike in mentions of overdose and death, despite not much evidence of an upward trend pre-pandemic. This highlights the role social media may be able to play as an early warning system: this spike presaged the very large spike in overdose deaths later observed in official death records, as detailed in the Introduction.

Of particular note is the steep drop in crime-related mentions (being arrested or busted) and money-related mentions. The money topic was intended to capture conversations relating to money and monetary stressors. While in early weeks many feared considerable economic hardship during the pandemic, the impact of generous social and governmental relief programs and the expansion of the social safety net enacted during the pandemic was actually to \emph{reduce} poverty, cutting poverty rates nearly in half compared to pre-pandemic levels \cite{wheaton_2021_2021}. 

\subsection{Geographic Analysis of Black-market Opiates Experience during the Covid-19 Pandemic}

Next, we present discussions by topic for users in our geographic cohort, split into users living in states with a Trump vote share (averaged between the 2016 and 2020 elections) over and under 50\%. We use this vote share variable as a  proxy for state government approaches towards managing the pandemic. States that had higher Trump vote shares are more likely to be led by Republicans, and to have reopened earlier, with their leaders placing greater emphasis on the economic and social harms of lockdowns.

\begin{figure*}[htbp]
\includegraphics[width=0.24\textwidth]{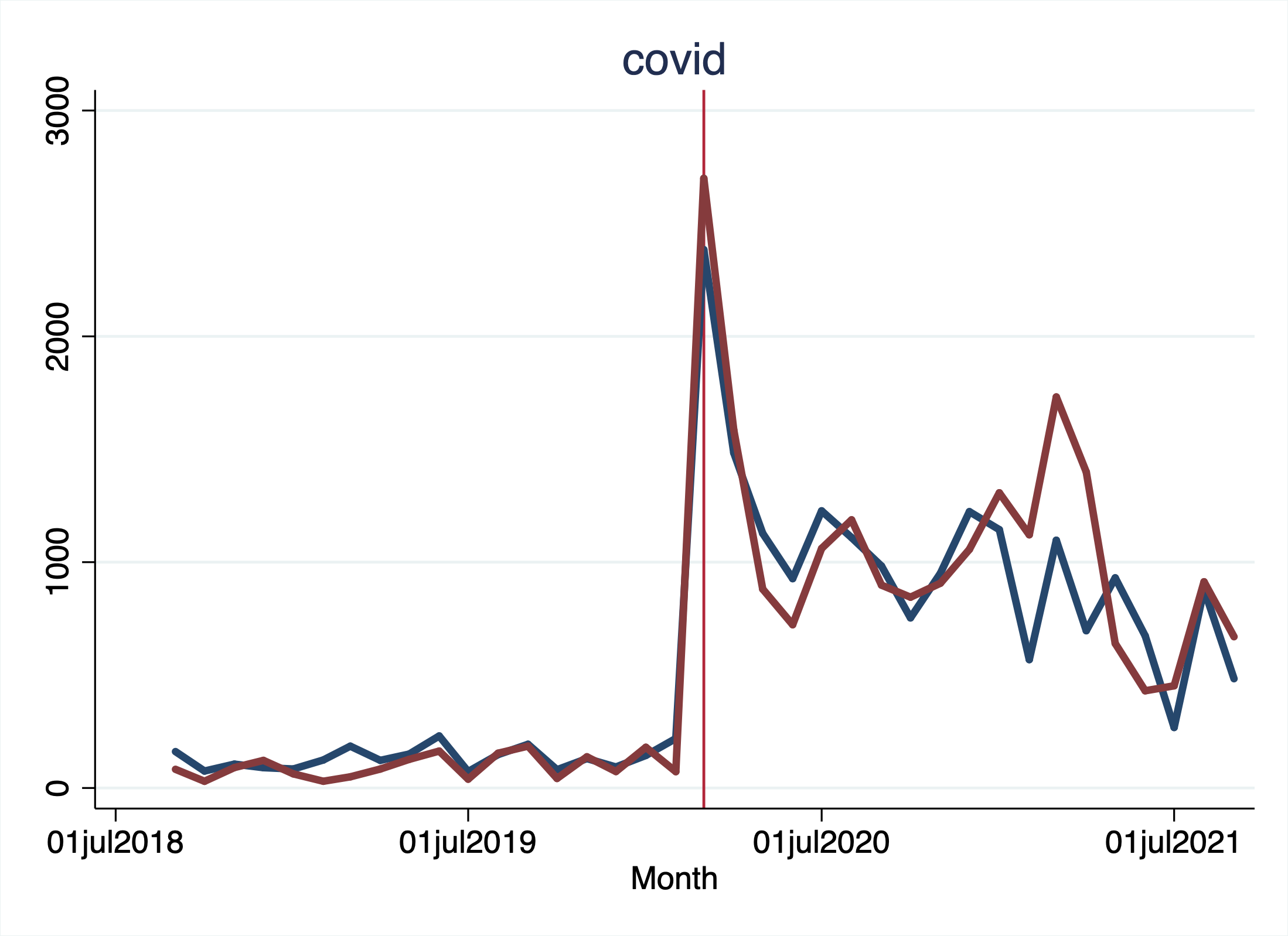}%
\includegraphics[width=0.24\textwidth]{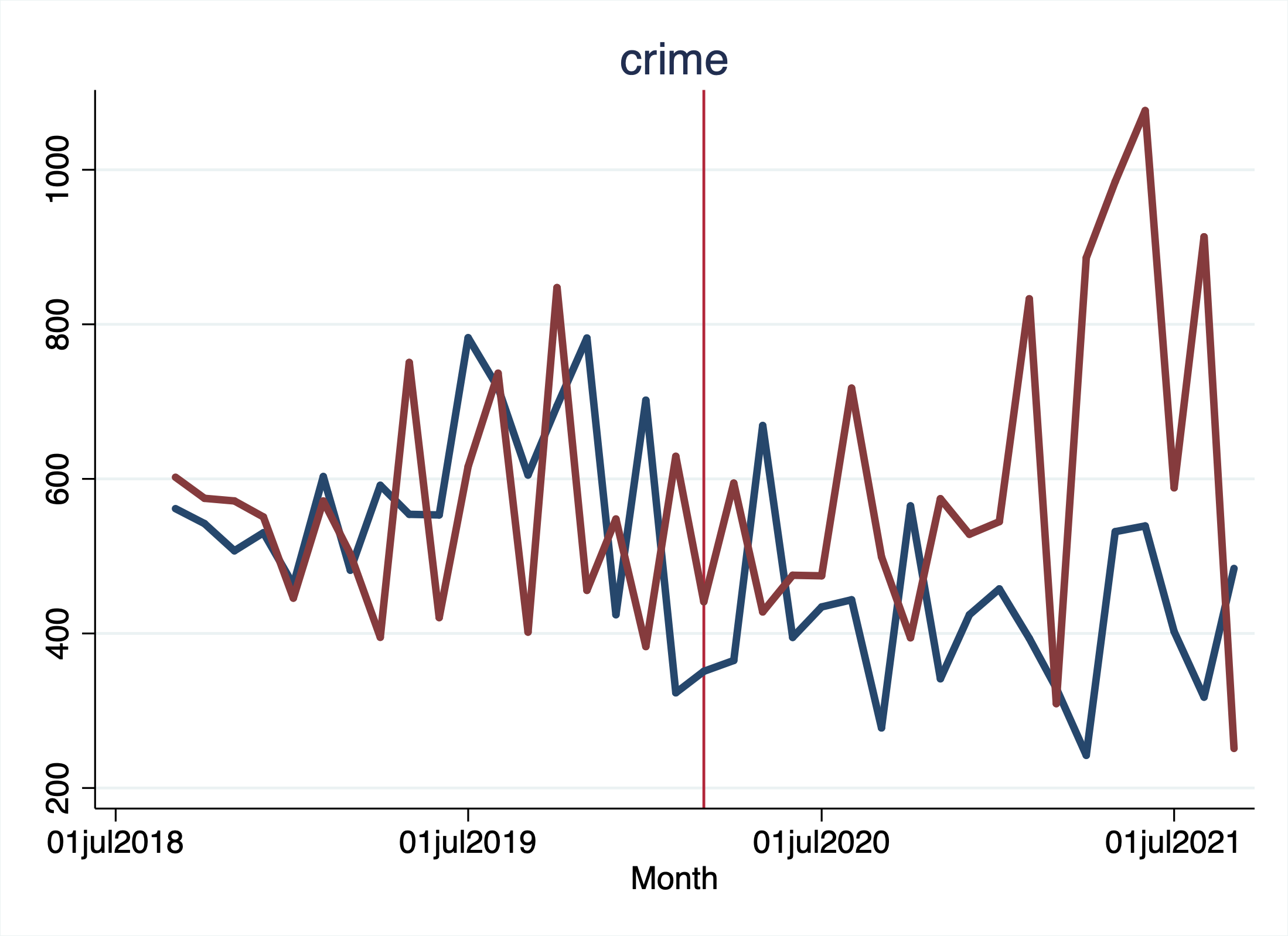}%
\includegraphics[width=0.24\textwidth]{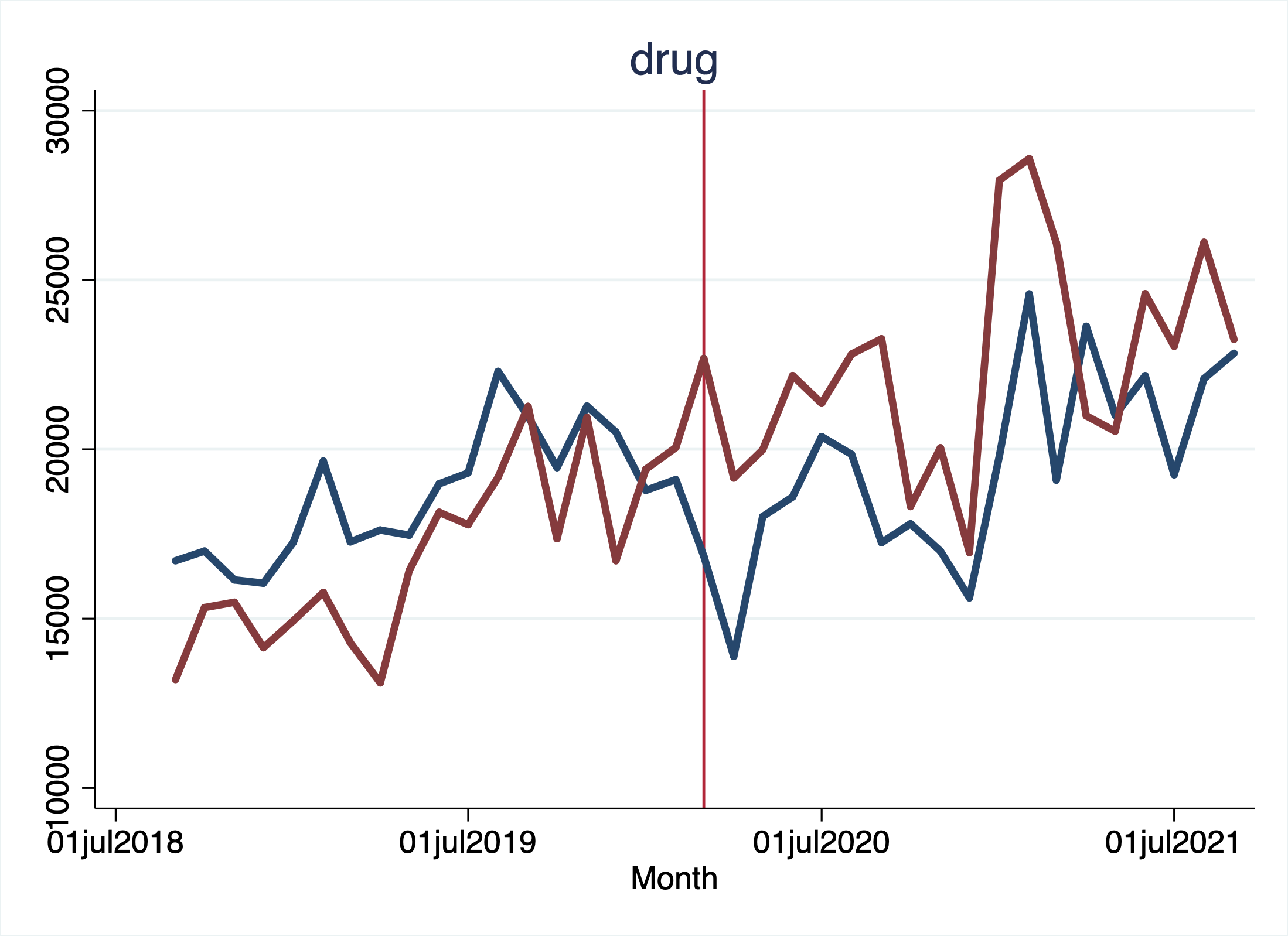}
\includegraphics[width=0.24\textwidth]{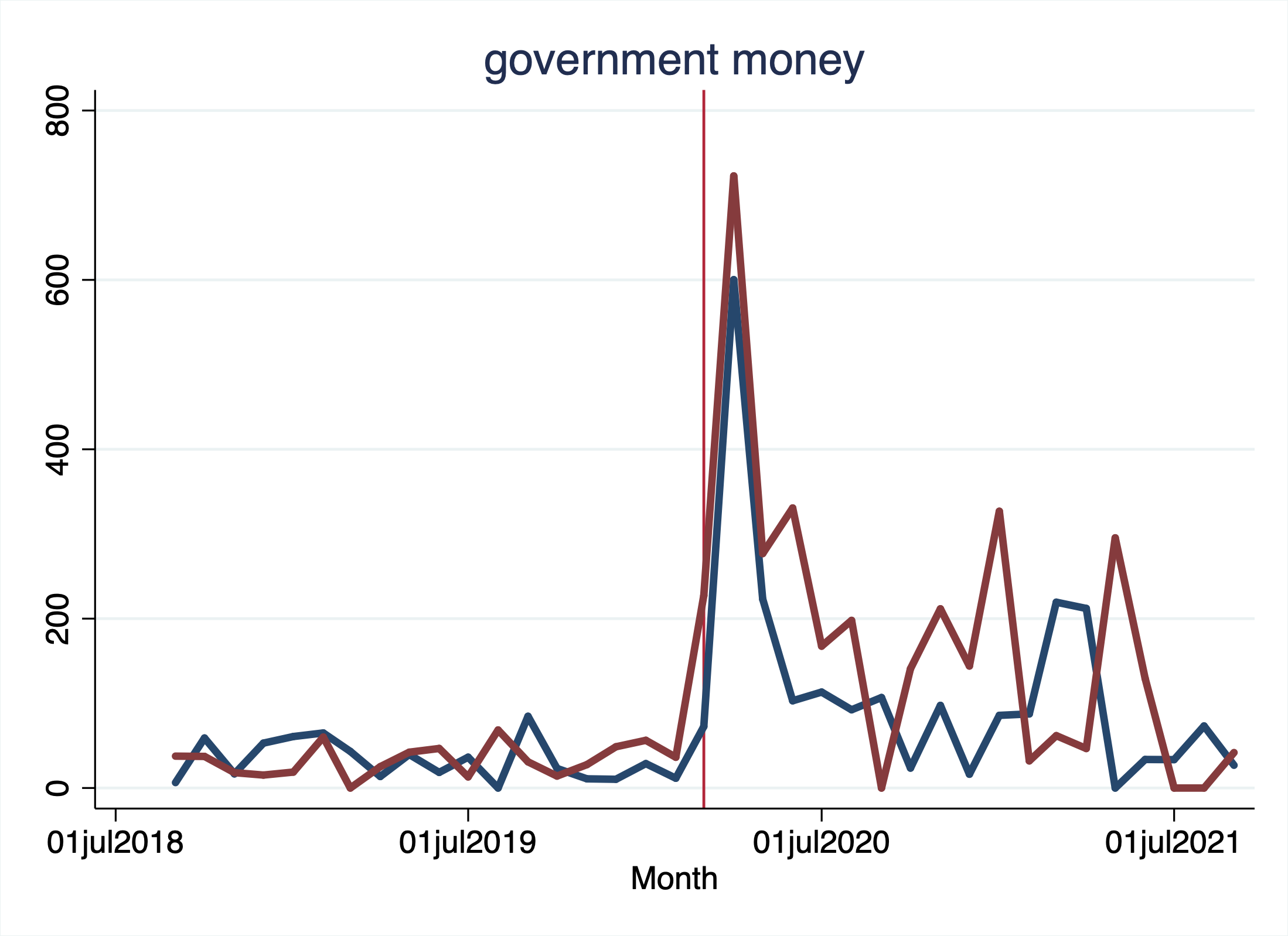}%
\includegraphics[width=0.24\textwidth]{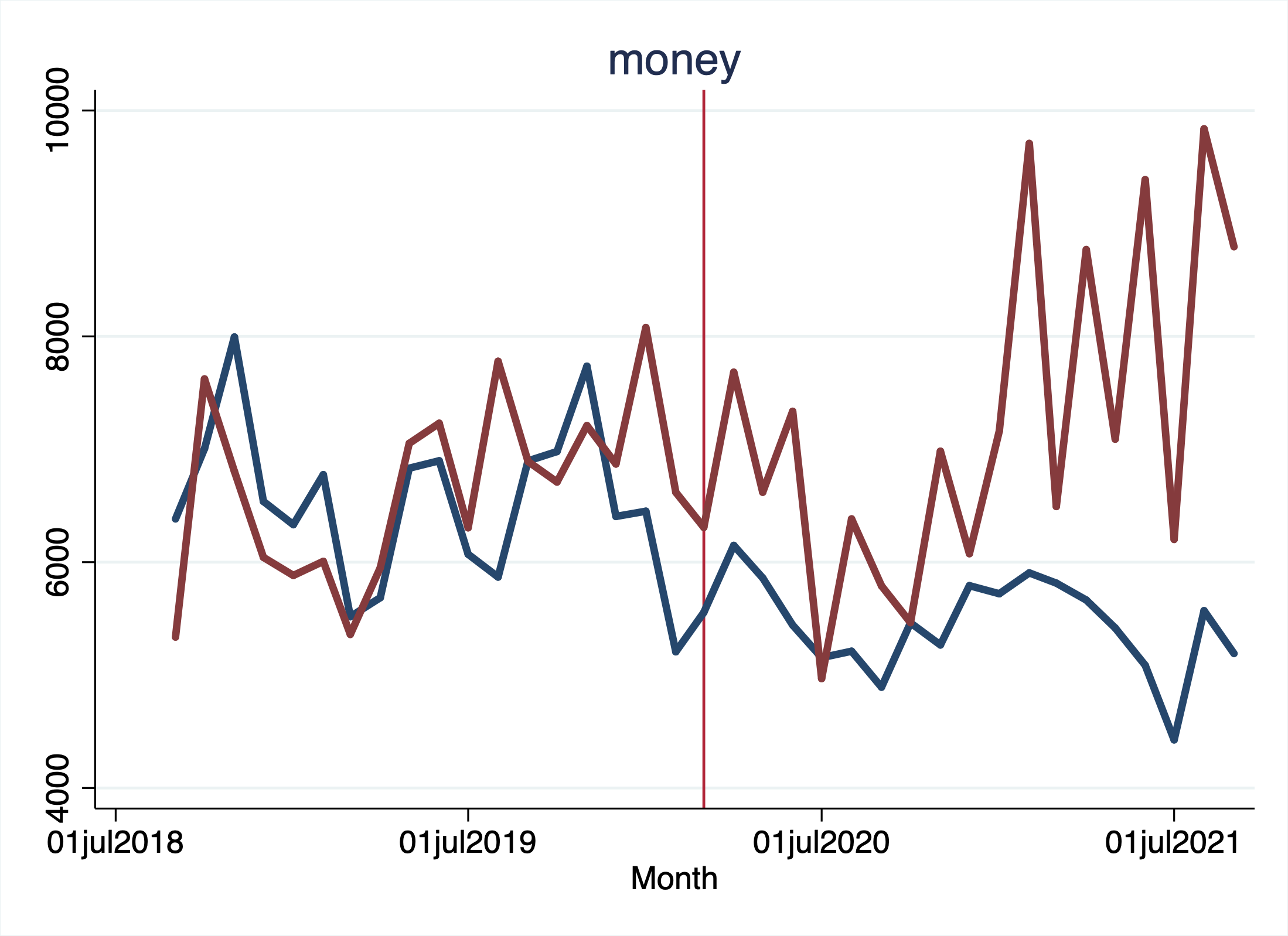}%
\includegraphics[width=0.24\textwidth]{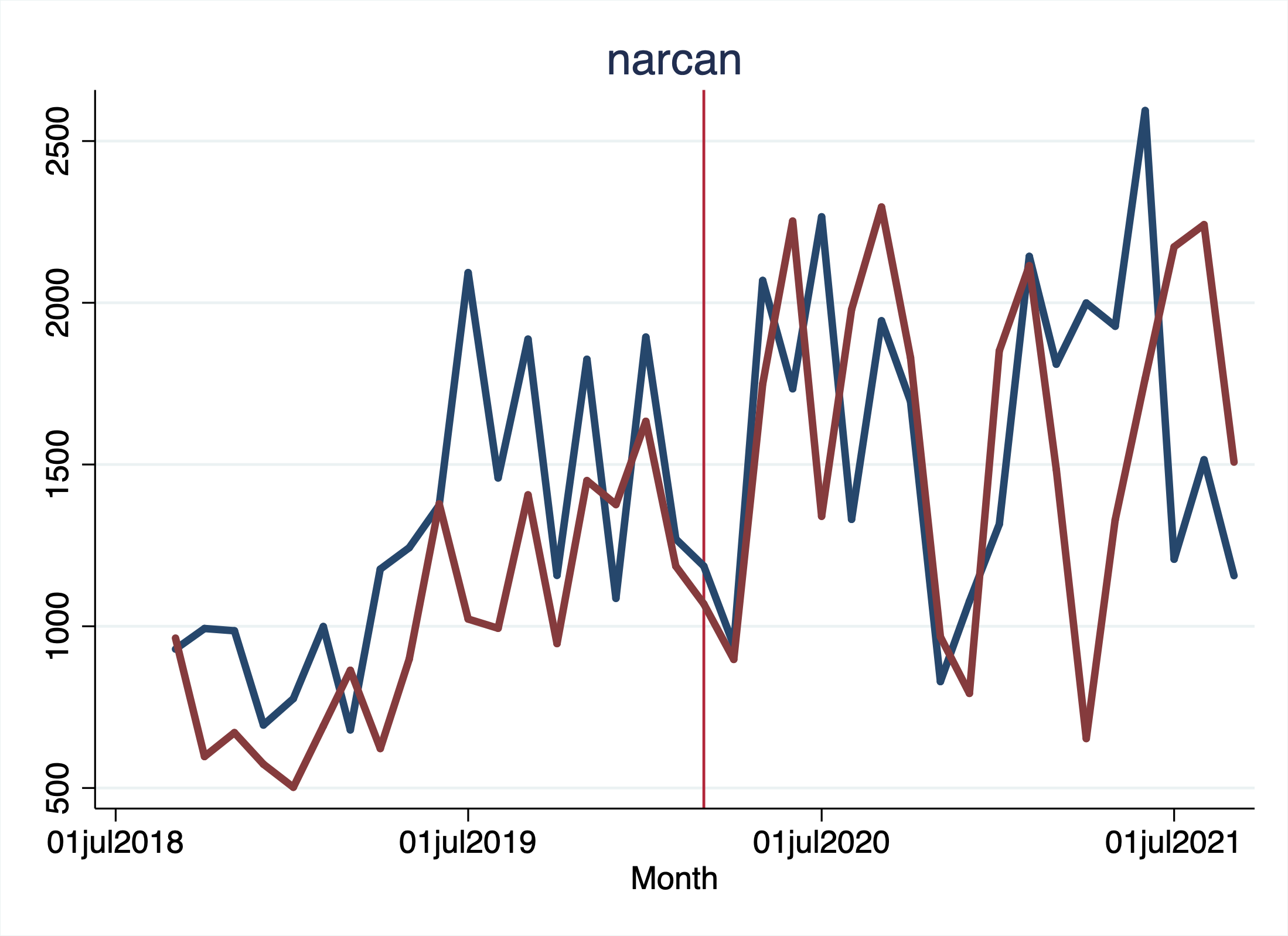}
\includegraphics[width=0.24\textwidth]{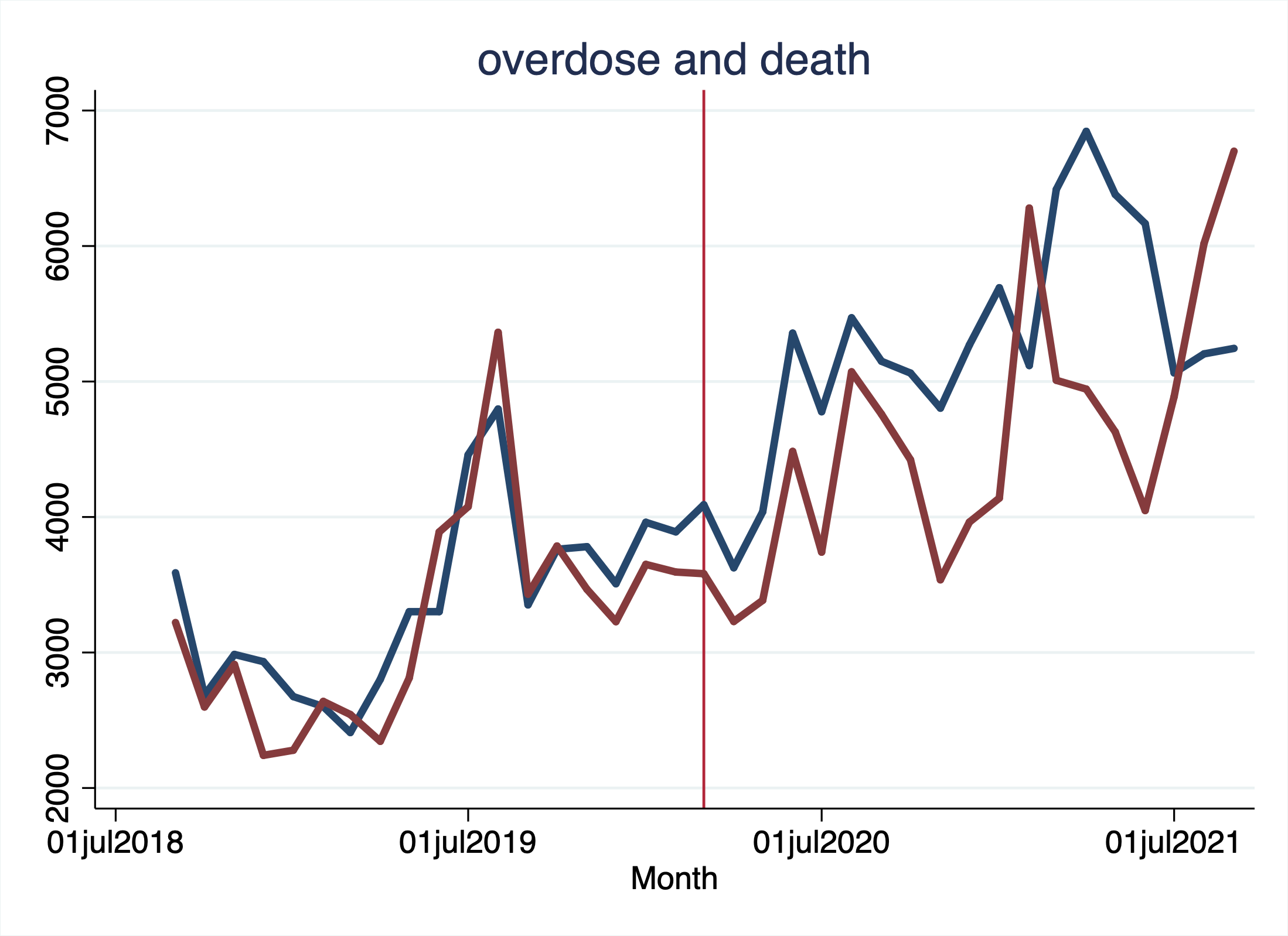}%
\includegraphics[width=0.24\textwidth]{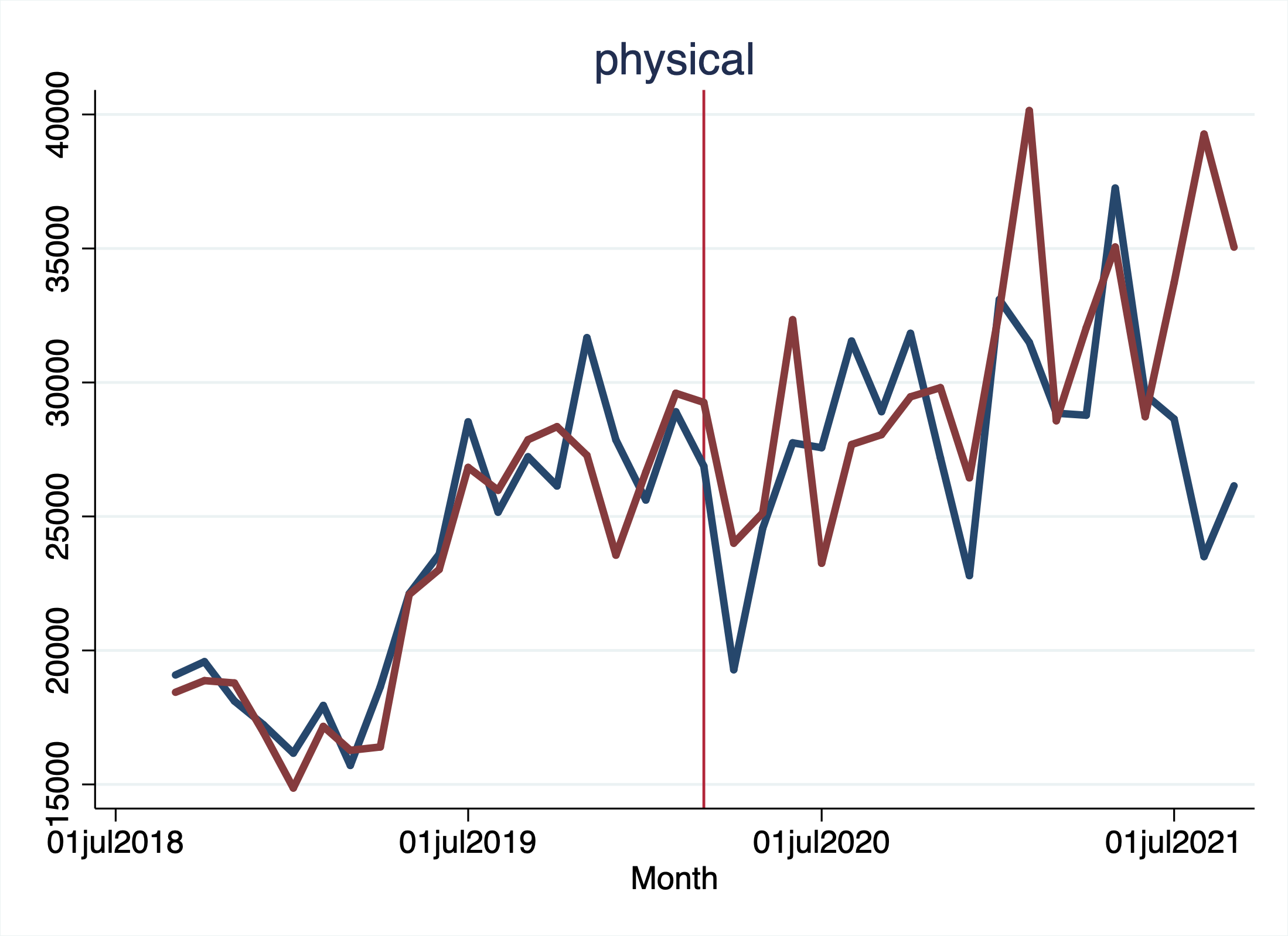}%
\includegraphics[width=0.24\textwidth]{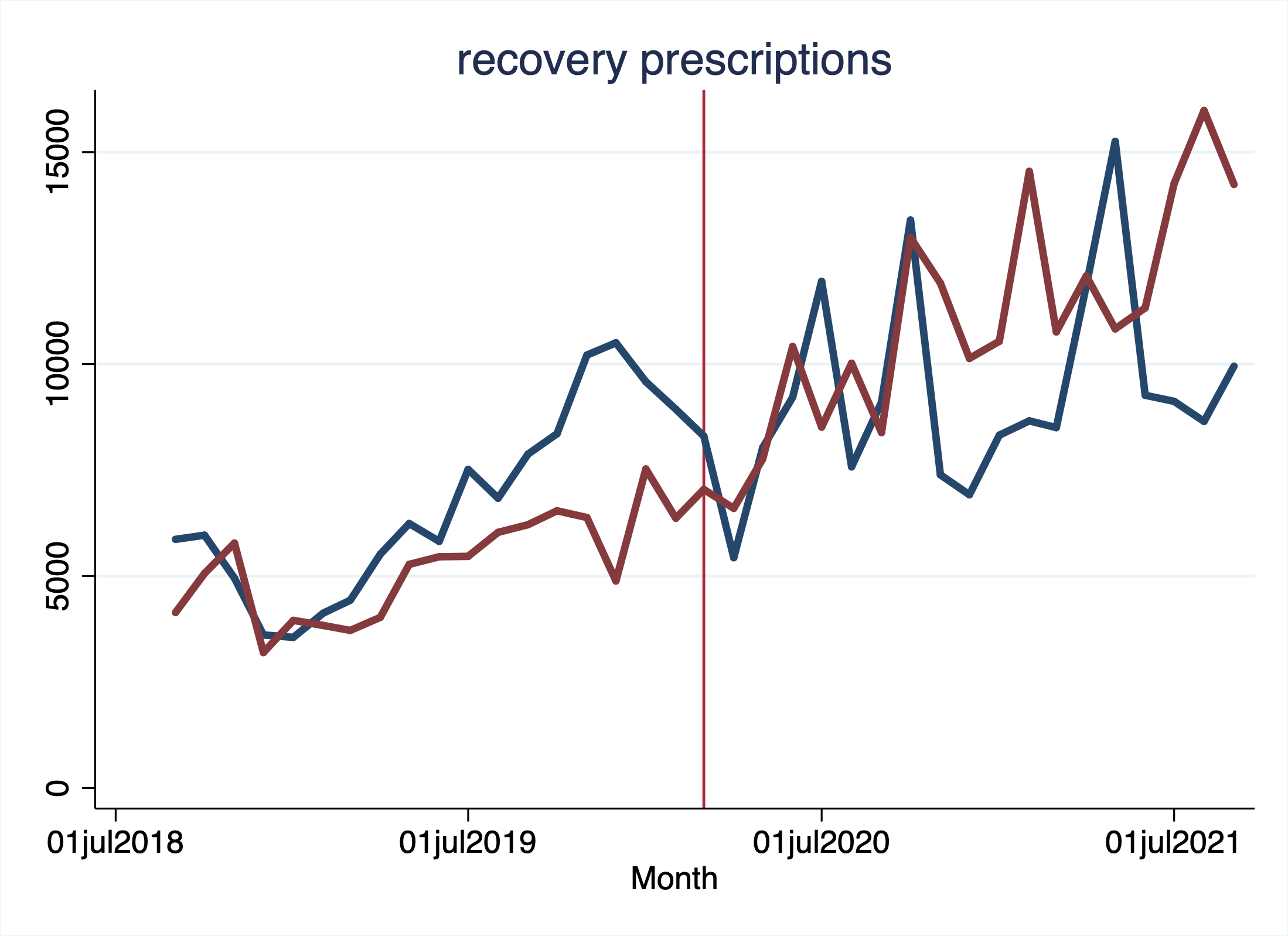}
\caption{Adjusted posting volume on geolocated user cohort for period of study analysis, September 2018 to September 21. Red lines are users in states with average Trump 2016 and 2020 vote share $> 0.5$ and blue lines are users in states with average Trump 2016 and 2020 vote share $<= 0.5$. Vertical red lines demarcate March 2020, the beginning of the pandemic}
\label{figvolumekeypolitics}
\end{figure*}

The results are presented in Figure \ref{figvolumekeypolitics}. The graphs depict that across some measures there are notable differences between users in states with a Trump vote share over 50\% (depicted with red lines, the ``red states'') compared to those below 50\% (depicted with blue lines, the ``blue states''). The most striking are the results on crime and money - while nationally mentions of these two topics fell, and they also fell in blue states throughout the time period of study, in red states they began to trend upwards in early-to-mid 2021. This corresponds with the timing of red states aggressively moving to reopen their economies; many ended the government's social support programs early to encourage people to return to work. Recent research has shown that as expanded benefits have expired, the poverty rate has ticked sharply upwards, reversing the reductions in poverty observed during the main period of the pandemic \cite{parolin_monthly_2020, jones_what_2021}. These national trends are clearly reflected in the experiences of black-market opiates users.  

\begin{table}
  \caption{Linear Regression Results for Topics}
\textbf{Panel A}
{
\def\sym#1{\ifmmode^{#1}\else\(^{#1}\)\fi}
\resizebox{\columnwidth}{!}{%
\begin{tabular}{l*{5}{c}}
\hline\hline
          &\multicolumn{1}{c}{(1)}&\multicolumn{1}{c}{(2)}&\multicolumn{1}{c}{(3)}&\multicolumn{1}{c}{(4)}&\multicolumn{1}{c}{(5)}\\
          &\multicolumn{1}{c}{covid}&\multicolumn{1}{c}{crime}&\multicolumn{1}{c}{drug}&\multicolumn{1}{c}{gov money}&\multicolumn{1}{c}{money}\\
\hline
Post-covid    &    861.4\sym{***}&   -159.7\sym{**} &    798.4         &    84.57\sym{*}  &  -1080.1\sym{***}\\
          &  (115.5)         &  (51.62)         &  (869.4)         &  (36.92)         &  (304.8)         \\
[1em]
$> 0.5$ Trump    &   -38.01         &   -23.10         &  -1795.8\sym{*}  &    0.859         &    120.9         \\
          &  (117.1)         &  (52.31)         &  (881.1)         &  (37.42)         &  (308.9)         \\
[1em]
Post-covid  &  122.1         &    210.0\sym{**} &   4858.6\sym{***}&    58.73         &   1640.8\sym{***}\\
 \& $> 0.5$ Trump          &  (163.4)         &  (73.00)         & (1229.5)         &  (52.21)         &  (431.1)         \\
[1em]
Constant    &    134.0         &    578.9\sym{***}&  18657.5\sym{***}&    32.32         &   6532.1\sym{***}\\
          &  (82.77)         &  (36.99)         &  (623.0)         &  (26.46)         &  (218.4)         \\
\hline\hline
\end{tabular}%
}
}

\vspace{2mm}

\textbf{Panel B}

{
\def\sym#1{\ifmmode^{#1}\else\(^{#1}\)\fi}
\resizebox{\columnwidth}{!}{%
\begin{tabular}{l*{4}{c}}
\hline\hline
          &\multicolumn{1}{c}{(6)}&\multicolumn{1}{c}{(7)}&\multicolumn{1}{c}{(8)}&\multicolumn{1}{c}{(9)}\\
          &\multicolumn{1}{c}{overdose \& death}&\multicolumn{1}{c}{physical}&\multicolumn{1}{c}{recovery rx}&\multicolumn{1}{c}{narcan}\\
\hline
Post-covid     &   1874.2\sym{***}&   5452.0\sym{***}&   2648.9\sym{***}&    366.6\sym{*}  \\
          &  (270.3)         & (1545.3)         &  (715.4)         &  (150.6)         \\
[1em]
$> 0.5$ Trump    &   -151.0         &   -573.3         &  -1447.6\sym{*}  &   -263.7         \\
          &  (274.0)         & (1566.1)         &  (725.0)         &  (152.6)         \\
[1em]
Post-covid&   -530.5         &   2946.4         &   3102.4\sym{**} &    240.3         \\
   \& $> 0.5$ Trump         &  (382.3)         & (2185.4)         & (1011.7)         &  (213.0)         \\
[1em]
Constant    &   3377.4\sym{***}&  22738.4\sym{***}&   6661.4\sym{***}&   1251.4\sym{***}\\
          &  (193.7)         & (1107.4)         &  (512.6)         &  (107.9)         \\
\hline\hline
\multicolumn{5}{l}{\footnotesize Standard errors in parentheses}\\
\multicolumn{5}{l}{\footnotesize \sym{*} \(p<0.05\), \sym{**} \(p<0.01\), \sym{***} \(p<0.001\)}\\
\end{tabular}%
}
}
\label{regtable}
\end{table}

However, the results are mixed regarding the impact of economic and social stressors on drug using behavior and overdoses. Drug mentions are up slightly in red states, but if anything, overdose mentions are up in blue states. We also present formal regression results estimated according to the following regression equation:

		\begin{align*} \text{Adjusted Mentions}_t = \alpha + \beta_1 \,\, 1(\text{Post-covid}_t) + \beta_2 \, \, 1(>\text{0.5 Trump}_{t}) \\ +  \beta_3 \,\, 1(\text{Post-covid}_t) \,*\, 1(>\text{0.5 Trump}_{t})  + \epsilon_{t}  \label{eq} \end{align*}

The results are presented in Table \ref{regtable}. While the mentions of crime, money, and drugs are statistically significantly higher in red states post-Covid, the mentions of overdoses are not statistically significantly higher in blue states. Mentions of recovery prescriptions (Medication Assisted Treatment), or drugs that can help a user with getting off of abusing opiates, \emph{were} higher in red states. 

Overall, the study of this user cohort suggests that overdoses rose despite the expansion of the national social safety net, which appears to have meaningfully reduced economic stressors. When the safety net has been reduced or removed in some geographic areas of the country, economic stressors have returned for users living in those areas, but again, this does not appear correlated to additional overdoses. The results paint a picture of the lives of opioid users during the Covid-19 crisis, but are also highly suggestive that more study is needed to understand sharply rising overdose mortality in this group.

\section{Conclusion}
This work combined several techniques into an approach for inferring the location of users on social media in an unsupervised manner.  This approach was then applied to over 100k Reddit users that have posted in opioid related subreddits.  Using the results of this location inference task, the temporal, geographic, and posting behavior of these users were analyzed in the context of lockdowns during the coronavirus pandemic.  


\begin{acks}
This project received support from the Crime and Justice Policy Lab at the University of Pennsylvania, which was funded by the Arnold Foundation to examine harm reduction around the opioid epidemic. We gratefully acknowledge research help from Ciara Tenney, Jackson Reimer and Anvita Pandit. 
\end{acks}

\clearpage 
\pagebreak

\bibliographystyle{ACM-Reference-Format}
\bibliography{drug-pricing-bibtex}


\clearpage 
\pagebreak
\appendix

\setcounter{table}{0}
\renewcommand{\thetable}{A\arabic{table}}

\section{Location Inference Preprocessing}

\subsection{Geopolitical Entity Filtering}\label{GPEfilter}
We hardcode a few Geopolitical Entities produce by SpaCy (Table \ref{gpe}) to intentionally not include in our geocoding process because we believe that they are primarily involved with non-location related discourse. 

\begin{table}
  \caption{Blocked Geopolitical Entities}
  \label{gpe}
  \begin{tabular}{c}
    \toprule
    Entity Name\\
    \midrule
    China\\
    Russia\\
    Turkey\\
    Op\\
  \bottomrule
\end{tabular}
\end{table}

\subsection{Large State Breakdowns}
Since we set the maximum distance for any two points within a location cluster to be roughly 100 miles, it is possible for several clusters to exist within larger states.  In an attempt to allow the mention of a state name to be present in all possible clusters within a larger state, we expand the name to be multiple regions within that state. Therefore when these regions get geocoded, they are mapped to multiple latitude/longitude pairs within the state and can then be captured by multiple clusters existing within that state, as shown in Table \ref{state}.

\begin{table}
  \caption{Large State Region Expansions}
  \label{state}
  \begin{tabular}{cc}
    \toprule
    State Name & Region/City Names\\
    \midrule
    California & Central, Southern, Northern\\
    Texas & El Paso, Houston, Dallas\\
    Florida & Tallahassee, Miami\\
    Alaska & Juneau, Anchorage, Fairbanks\\
  \bottomrule
\end{tabular}
\end{table}

\subsection{Location Aliases}
\label{aliases}

We noticed a few location nicknames that were not being properly picked up by the named entity recognition tool in SpaCy, so we hardcoded aliases for these nicknames to more proper names to allow them to be detected.  These mappings are shown in Table \ref{aliastable}.

\begin{table}
  \caption{Location Nickname Mappings}
  \label{aliastable}
  \begin{tabular}{cc}
    \toprule
    Nickname & Proper Name\\
    \midrule
    Vegas & Las Vegas\\
    NYC & New York City\\
    L.A. & Los Angeles\\
  \bottomrule
\end{tabular}
\end{table}

\subsection*{}

\end{document}